\documentclass[preprint,12pt]{elsarticle}

\usepackage{amssymb}
\usepackage{amsmath}

\usepackage{lineno}

\usepackage{color}

\usepackage{multirow}
\usepackage{booktabs}

\usepackage{subcaption}
\usepackage{caption}
\usepackage{float}

\graphicspath{ {./images/} }

\newcommand{\REV}[1]{{\color{black}{#1}}}
\newcommand{\CORR}[1]{{\color{black}{#1}}}

\newcommand{\bn}{\boldsymbol}
\newcommand{\argmin}{\operatornamewithlimits{argmin}}

\renewcommand{\Re}{\operatorname{real}}

\usepackage{doi}

\journal{Computers in Biology and Medicine}

\begin{document}

\begin{frontmatter}



\title{A Weighted Hankel Approach and Cramér-Rao Bound Analysis for Quantitative Acoustic Microscopy Imaging}

\author[label1]{Lorena Leon}
\author[label2]{Jonathan Mamou}
\author[label3]{Denis Kouamé}
\author[label1]{Adrian Basarab}

\affiliation[label1]{organization={Univ Lyon, INSA-Lyon, Université Claude Bernard Lyon 1, CNRS,
Inserm, CREATIS UMR 5220, U1294},
            city={Villeurbanne},
            postcode={9100},
            country={France}}

\affiliation[label2]{organization={Department of Radiology,Weill Cornell Medicine},
            city={New York City},
            postcode={10022},
            state={NY},
            country={USA}}

\affiliation[label3]{organization={IRIT UMR CNRS 5505, Université Paul Sabatier Toulouse 3},
            city={Toulouse},
            postcode={31062},
            country={France}}



\begin{abstract}
Quantitative acoustic microscopy (QAM) is a cutting-edge imaging modality that leverages very high-frequency ultrasound to characterize the acoustic and mechanical properties of biological tissues at microscopic resolutions. Radio-frequency signals are digitized and processed to yield two-dimensional maps. This paper introduces a weighted Hankel-based spectral method with a reweighting strategy to enhance robustness with regard to noise and reduce unreliable acoustic parameter estimates. Additionally, we derive, for the first time in QAM, Cramér-Rao bounds to establish theoretical performance benchmarks for acoustic parameter estimation. Simulations and experimental results demonstrate that the proposed method consistently outperform standard autoregressive approach under challenging conditions. These advancements promise to improve the accuracy and reliability of tissue characterization, enhancing the potential of QAM for biomedical applications.
\end{abstract}







\begin{keyword}



Hankel spectral method \sep Cramér-Rao bound \sep Biomedical imaging \sep Quantitative acoustic microscopy

\end{keyword}

\end{frontmatter}




\section{Introduction} \label{sec:introduction}

Quantitative acoustic microscopy (QAM) employs ultrasound waves with frequencies exceeding 100~MHz to scan thin (5-14~$\mu$m thick) \emph{ex vivo} tissue sections affixed to microscopy slides, using filtered water as the coupling medium~\cite{1975_Lemons,Hoerig2023}. QAM is closely related to scanning acoustic microscopy (SAM), a well-established technique in nondestructive testing~\cite{1985introduction,Lemons1974} and in the study of hard biological samples (e.g., bone~\cite{Raum2008}). Although SAM has also been applied to soft tissue imaging~\cite{Litniewski1990,Miura2013}, providing insights into acoustic properties at frequencies near 100~MHz, its primary outputs are typically qualitative contrast maps, such as A-scans, B-scans, and C-scans.

In contrast to SAM, QAM produces two-dimensional (2D) quantitative maps of acoustic properties, including speed of sound, acoustic impedance, and attenuation, as well as mechanical properties such as bulk modulus, density, and compressibility, achieving spatial resolutions close to the acoustic wavelength (e.g., finer than 8~$\mu$m for a 250~MHz QAM system~\cite{Rohrbach2015retine}). {This transition from qualitative to quantitative imaging represents a significant advance, establishing QAM as a powerful and unique modality for biomedical research, offering} complementary advantages over conventional microscopy modalities such as optical microscopy, electron microscopy, and histological imaging for detailed tissue characterization~\cite{Hildebrand1981,Ogawa19}. 
Its versatility has been demonstrated in a wide range of applications, including corneal imaging~\cite{Rohrbach2018cornea}, retinal studies~\cite{Marmor77,Rohrbach2015retine}, liver tissue analysis~\cite{Irie2016}, investigations of cerebellar folding~\cite{Lawton2019}, assessments of biomechanical changes in myopic eyes~\cite{Rohrbach2015,HOANG2019107739}, studies of cancerous lymph nodes~\cite{Mamou2015}, {renal fibrosis~\cite{Ito2025}, cutaneous melanoma~\cite{Arakawa_2018}},  and analyses of living cells~\cite{Hildebrand1981,BRIGGS1993,Weiss2007,Miura2015}. {Another study in the area highlights the relevance of acoustic parameters for tissue characterization and their dependence on spatial resolution~\cite{Ogawa_2020}.}



Data acquired by QAM instruments consist of radio frequency (RF) signals containing two main reflections originating from the water-sample and sample-glass interfaces. These RF signals are then processed offline using model-based approaches to estimate acoustic parameters at each scanned position. While time domain methods~\cite{BRIGGS1993} have been explored, frequency domain techniques have proven advantageous, particularly when the reflections overlap in time~\cite{Hozumi2004}. The current QAM methodology, introduced in~\cite{Rohrbach2018}, relies on an autoregressive (AR) frequency domain model combined with a spectral estimation algorithm inspired by the classical Prony method~\cite{prony} and the annihilating filter~\cite{jin2017mri}.

Despite the advantages of the AR-based method, QAM continues to face challenges in decomposing signals into primary reflection components and obtaining reliable estimates. These challenges become particularly critical when dealing with high noise levels, thin tissue samples (i.e., $\sim$ acoustic wavelength), small impedance contrast, or large tissue attenuation. These difficulties limit the full potential of QAM for high-resolution tissue characterization.


The goal of this work is to propose a robust spectral-based QAM framework to address the aforementioned limitations in acoustic parameter estimation. To that end, the QAM measurement forward model and image formation process are briefly recalled in Section~\ref{sec: model}.

The first contribution, presented in Section~\ref{sec:HK}, is the introduction of a weighted Hankel-based method that builds on Hankel matrix theory, as inspired by~\cite{Andersson2014}, to address the spectral estimation problem. 

\REV{Related approaches that leverage structured low-rank modeling have been successfully applied to other imaging modalities, including computed tomography~\cite{Lobos2021}, spectroscopy~\cite{Qiu2020}, and image imputation tasks~\cite{Jin15}. Structured low-rank methods have also been widely used in magnetic resonance imaging (MRI) reconstruction, where they have been employed to model spatially limited and phase-constrained images~\cite{Shin2013, Haldar2014}, as well as transform-sparse images~\cite{Ongie2016, Jin2016}. For a comprehensive overview of structured low-rank modeling in MRI, see the review article~\cite{Haldar2020}.}

\REV{Our approach shares several conceptual similarities with prior structured low-rank methods, although the specific matrix structures and application contexts differ. For instance, in MRI, k-space data are typically modeled using block-Hankel matrices, while in our case, a single Hankel matrix is constructed from the spectral data. Unlike in MRI where low-rank modeling is often central to image reconstruction, in QAM, spectral reconstruction via low-rank approximation primarily serves as an initial denoising step. Subsequent analysis requires identifying the two dominant signal components and estimating their parameters using an ESPRIT-type algorithm applied directly to the Hankel matrix formed from the denoised spectrum, rather than using a traditional covariance matrix, as is common in standard ESPRIT~\cite{Esprit}.}

\REV{In contrast to fast MRI, where spectral data are frequently undersampled and require sophisticated reconstruction techniques, in QAM we work with a fully sampled spectrum. Additionally, our setting is inherently one-dimensional RF signals, whereas MRI typically involves two or three dimensional Fourier representations with geometric (gradient) information. These distinctions motivate our adaptations of existing structured low-rank approaches to suit QAM data.}

\REV{A key contribution of this work is the introduction of a reweighting strategy designed to better manage noise-corrupted components in the normalized spectrum, thereby extending our earlier approach~\cite{leon2024isbi}. This is motivated by the trade-off between including a sufficient number of relevant frequency components and excluding corrupted ones, an issue that is particularly sensitive to the bandwidth selection.}

\REV{While our approach focuses on reweighting the data fidelity term, it is worth noting that iterative reweighting strategies have also been proposed for regularizing the Hankel matrix itself, particularly in the context of low-rank priors in the Fourier domain for MRI reconstruction~\cite{Ongie2017}, and more generally in the context of low-rank matrix completion~\cite{Fornasier11, Mohan2012}. However, to the best of our knowledge, our work represents the first adaptation of such techniques to the specific context and signal model of QAM data.}

\REV{Another significant contribution of this work is the derivation of the Cramér-Rao bounds in the context of QAM, which is presented for the first time, as detailed in Section~\ref{sec:CRB}. These bounds set theoretical limits on the performance of estimators, enabling comparison of different estimation methods. They reveal the inherent challenges of estimating specific parameters under different conditions. They also facilitate the validation of physical modeling assumptions and help define experimental setups that promote reliable parameter estimation.}

In addition, the proposed methods are compared against the standard AR model through extensive simulations and experimental data obtained from soft tissue using a QAM instrument operating at 500~MHz, as detailed in Section~\ref{sec:material}. The performance results, presented in Section~\ref{sec:results}, demonstrate that the proposed methodology significantly reduces unreliable estimates while maintaining or improving estimation accuracy under challenging conditions.

Finally, Section~\ref{sec:conc} discusses the findings and outlines potential directions for future research.





\section{Signal Model in QAM}\label{sec: model}

\begin{figure}[tb]
\centering
\begin{subfigure}{0.4\linewidth}
\centering
\caption{Experimental QAM approach}
\includegraphics[width=0.8\linewidth, trim=0 460 850 0,clip]{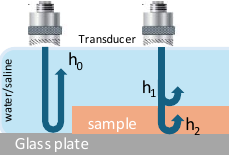} 
\end{subfigure}

\vspace{0.01em}
        \centering
        \begin{subfigure}{0.48\linewidth}
            \caption{Reference signal}
            \centering
            \includegraphics[width=\linewidth, trim=10 10 25 10,clip]{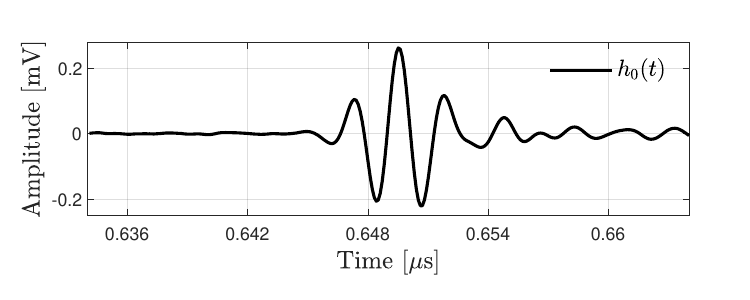}            
        \end{subfigure}
        \hfill
        \begin{subfigure}{0.48\linewidth}
            \caption{QAM signal}
            \centering
            \includegraphics[width=\linewidth, trim=10 10 25 10,clip]{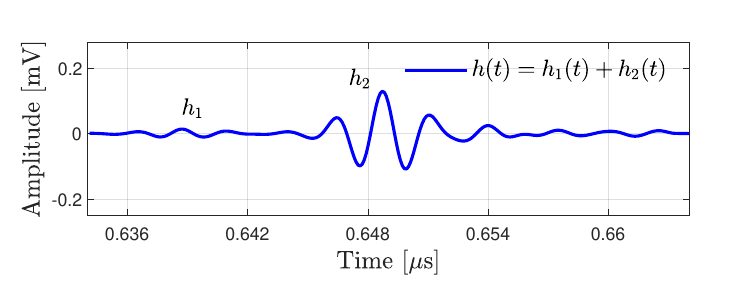}   
        \end{subfigure}
        \vspace{0.5em}
        \begin{subfigure}{0.48\linewidth}
        \caption{Water-tissue echo}
            \centering
            \includegraphics[width=\linewidth, trim=10 10 25 10,clip]{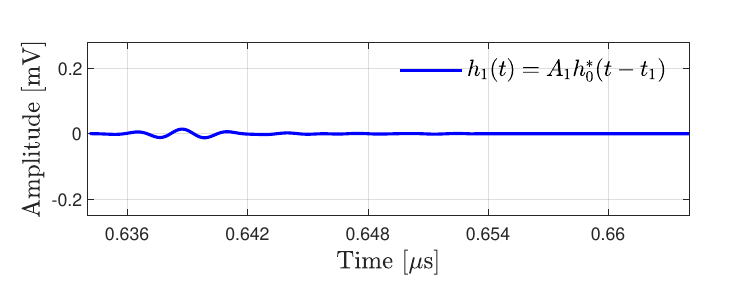}          
        \end{subfigure}
        \hfill
        \begin{subfigure}{0.48\linewidth}
        \caption{Tissue-glass echo}
            \centering
           \includegraphics[width=\linewidth, trim=10 10 25 10,clip]{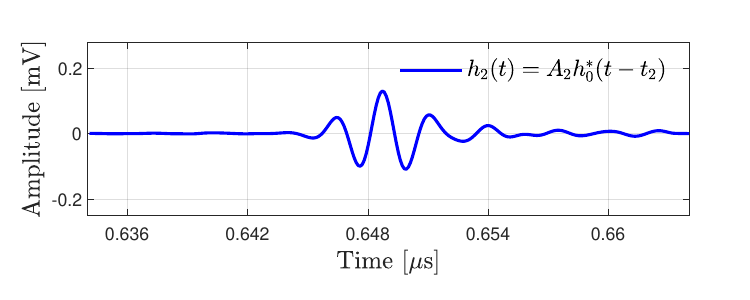}   
        \end{subfigure}
    \caption{{(a) Diagram of the experimental QAM approach. The reference signal $h_0(t)$, shown in (b), is obtained from a single reflection at the glass-water interface in the absence of tissue. The measured signal $h(t)$ in (c) consists of two components, $h_1(t)$ and $h_2(t)$, associated with reflections at the water-tissue and tissue-glass interfaces, respectively. The echo signals $h_1(t)$ and $h_2(t)$ are shown in (d) and (e).
}}
    \label{fig:qam_principle}
\end{figure}

In QAM, an RF signal acquired in a tissue-free location results from a single reflection from the glass-water interface and serves as a reference signal, denoted by $h_0(t)$, where $t$ is the time (in seconds). 
On the other hand, a signal acquired at a tissue location, denoted by $h(t)$, is composed of two main reflections: one from the water-tissue interface $h_{1}(t)$ and the other from the tissue-glass interface $h_{2}(t)$. Fig.~\ref{fig:qam_principle}  illustrates the measurement principle in a tissue location (i.e.,~$h_{1}$ and $h_{2}$) and a tissue-free location (i.e.,~$h_0$).

In this context, the estimation of the acoustic properties of the sample relies on the accurate identification and characterization of~$h_{1}(t)$ and~$h_{2}(t)$ in terms of their amplitude, frequency attenuation, and time shift relative to ~$h_0(t)$. This section provides an overview of the signal model, the estimation problem, and the image formation process involved in QAM technology.


\subsection{From time to frequency domain}
{We consider single layer samples, however t}o account for the complex interactions of the acoustic waves within the tissue, including potential scattering effects and multiple reflections, we consider 
\begin{align}\label{timemodel}
h(t) = \sum_{p=1}^P h_p(t) \mbox{ with } P\geq 2,
\end{align}
where $P$, known as the model order, represents the number of reflections present in the acquired signal~\cite{Rohrbach2018}. At this stage, we assume $P$ to be given, and the choice of its value is discussed in Section~\ref{prepo}. 

Each echo $h_p$ in~\eqref{timemodel} can be modeled as a phase-shifted and amplitude-modified version of the reference signal $h_0$, i.e.,
\begin{align}\label{timemodel2}
h_p(t) = A_p h^{{*}}_0(t - t_p),
\end{align}
where $A_p$ is the real amplitude, $t_p$ is the time of flight (TOF) and {$*$} symbolizes an additional frequency-dependent attenuation effect due to the round-trip propagation inside the sample. {For consistency in the notation, the signals related to the water-tissue and tissue-glass interfaces remain denoted by the subscripts $p\!=\!1$ and $p\!=\!2$.}

Parameter estimation based on~\eqref{timemodel}-\eqref{timemodel2} becomes particularly challenging when echo signals overlap in time, as this complicates their separation. Furthermore, identifying the tissue-glass reflection becomes problematic when the sample is very thin or its impedance is close to that of the water used as a coupling medium. To overcome these challenges, the transition to the frequency domain has proven to be an effective approach~\cite{Rohrbach2018}.

{Taking the Fourier transform (FT) of~\eqref{timemodel}-\eqref{timemodel2}, and assuming frequency-dependent attenuation is linear with both frequency and propagation distance, we obtain
\begin{equation}\label{fourier_cont}
    \mathcal{F}_{h}(f)=\operatorname{FT}(h(t))=\mathcal{F}_{h_0}(f) \sum_{p=1}^P {A_p} \exp {(2\pi f[-{\beta_p}-\mathrm{i} {\Delta t_p}])}.
\end{equation}
Here, the ``$*$” were dropped to introduce the explicit frequency-dependent attenuation coefficients $\beta_p$ (in Np/MHz). It is also assumed that $\beta_1 = 0$ Np/MHz, implying negligible attenuation for the reflection from the water-tissue interface. In~\eqref{fourier_cont}, the relative time shift of the $p$th echo compared to the reference signal is given by $\Delta t_p\!=\!t_p-t_0$.}

{The ratio between $\mathcal{F}_{h}$ and $\mathcal{F}_{h_0}$
at discrete frequencies $k\Delta f$ can be written as
\begin{align}\label{fourier}
\dfrac{\mathcal{F}_h(k\Delta f)}{\mathcal{F}_{h_0}(k\Delta f)}\! =\!\sum_{p=1}^P {A_p} [\exp {(2\pi\Delta f[-{\beta_p}-\mathrm{i} {\Delta t_p}])}]^k,
\end{align}
where $\Delta f\!=\!f_s/(2M)$ denotes the frequency step size, $M$ represents the number of points in the sampled signal (considering no zero-padding),
and $f_s$ is the sampling rate.} 

The normalization (i.e., division) in~\eqref{fourier} is well-defined where $\mathcal{F}_{h_0}$ is non-zero, which is guaranteed for frequencies within the bandwidth of the transducer where sufficiently high signal-to-noise ratio can be achieved. At this stage, we assume the bandwidth is known, with $k_{\text{min}}$ and $k_{\text{max}}$ defining the frequency limits as $k_{\text{min}}\Delta f$ and $k_{\text{max}}\Delta f$, respectively. For simplicity, we assume $k_{\text{max}} - k_{\text{min}} = 2N$, resulting in a total of $2N+1$ frequencies. The choice of bandwidth is further discussed in Section~\ref{prepo}.

A noisy observation of the spectrum defined in~\eqref{fourier} can be seen as a special case of the following general parametric model
\begin{equation}\label{general process}
\begin{aligned}
x_{\kappa}&= \sum_{p=1}^Pa_p\lambda_p^\kappa+\epsilon_\kappa,
\end{aligned}
\end{equation}
with $\lambda_p\!=\!\exp{(2\pi(\gamma_p+\mathrm{i}\nu_p))}$, where $\gamma_p$ and $\nu_p$ are the damping and the frequency parameters; and $\epsilon_\kappa$ is a complex additive noise. Furthermore, $a_p\!=\!A_p\exp{(\mathrm{i} b_p)}$ represents the complex-valued amplitude, where $A_p\!=\!|a_p|$ is the real-valued amplitude, and $b_p\!=\!\angle a_p$ is the initial phase. By defining $\kappa\!=\!k/(2M)$, a noisy observation of~\eqref{fourier} and model~\eqref{general process} are equivalent if $b_p\!=\!0$, $\gamma_p\!=\!-\beta_p f_s$, $\nu_p\!=\!-\Delta t_pf_s$ for all $p\!=\!1,\ldots,P$.

The model~\eqref{general process} arises in various applications, such as astronomy, radar, communications, economics, medical imaging, and spectroscopy, to name but a few (see, e.g.,~\cite{stoica2005spectral} and the references therein). Unlike~\cite{Rohrbach2018} and~\cite{leon2024isbi}, this work opts to present the spectral model in its general form~\eqref{general process} instead of~\eqref{fourier} to facilitate the derivation of the theoretical bounds in Section~\ref{sec:CRB}. 



\subsection{Acoustic parameter estimation and image formation}\label{acust}

Given $\{\lambda_p\}_{p=1}^{P}$ and $\{a_p\}_{p=1}^{P}$, the two leading reflections can be determined by reconstructing the $P$ pulses and computing the amplitude of the maximum of the Hilbert-transformed signals. The two pulses with the largest amplitudes correspond to the desired components related to the water-tissue and tissue-glass interfaces, which are denoted with the subscripts $p\!=\!1$ and $p\!=\!2$, without loss of generality. 
Then, the signal corresponding to water-sample interface is identified as the one with the smallest TOF ($t_1 \!<\! t_2$), as it occurs before the sample-glass interface~\cite{Rohrbach2018}.

Following the derivations in~\cite{Rohrbach2018} and using the parameter notation from~\eqref{general process}, the speed of sound $c$ (in m/s) and the acoustic impedance $Z$ (in MRayl) can be computed as
\begin{align}\label{eq: c and Z} 
c\!&=\!c_{\text{w}}\dfrac{\nu_{1}}{\nu_{1}-\nu_{2}}, \quad Z\!=\!Z_{\text{w}}\dfrac{1+\frac{A_{1}}{R_{\text{wg}}}}{1-\frac{A_{1}}{R_{\text{wg}}}}, 
\end{align}
where $c_{\text{w}}$ and $Z_{\text{w}}$ are the known speed of sound and acoustic impedance in the water, and $R_{\text{wg}}$ is the known pressure reflection coefficient between water and glass. 

The tissue attenuation $\alpha\!=\!\beta_2/(2d)$ (in dB/MHz/cm) 
quantifies the signal attenuation during its round-trip travel through tissue with a thickness $d$ (in $\mu$m). Their values are obtained as follows,
\begin{align}\label{eq d and alpha}
d\!&=\!\dfrac{c_{\text{w}}}{2}\dfrac{\nu_{1}}{f_s}, \quad \alpha\!=\!-\dfrac{\gamma_2}{c_w \nu_{1} }.
\end{align}
It is important to note that a linear attenuation model is assumed to be valid within the specified bandwidth; for a nonlinear attenuation model, refer to~\cite{Rohrbach2018}.

Finally, a 2D map of each acoustic tissue property is obtained by combining the estimates obtained by applying the parameter-estimation algorithm to all scanned locations.


\section{Weighted Hankel-based parameter estimation} \label{sec:HK}

The accurate estimation of $c$, $Z$, $\alpha$ and $d$ depends fundamentally on the precise determination of the spectral parameters $(\lambda_1, a_1)$ and $(\lambda_2, a_2)$. Once $\lambda_p$ is known, the amplitudes $a_p$ can be recovered efficiently using a least squares (LS) approach. However, accurate estimation of $\lambda_p$ can be challenging. 

To overcome the limitations of the standard AR-based framework for QAM~\cite{Rohrbach2018}, we previously introduced a Hankel-based spectral estimation framework for QAM image formation~\cite{leon2024isbi}. This method builds upon the spectral estimation technique originally presented in~\cite{Andersson2014}. \CORR{In contrast to our previous approach~\cite{leon2024isbi}, an iterative reweighting strategy is incorporated, to further improve robustness by mitigating the influence of outliers in the spectral data and minimize unreliable acoustic parameter estimates.}

\subsection{Hankel-based problem formulation}

The Hankel-based spectral method applies to the parametric model \eqref{general process} and seeks to approximate the vector $\bn{x}\!=\!(x_{\kappa_1},\ldots,x_{\kappa_{2N+1}})$ by $\bn{g}\!=\!(g_{\kappa_1},\ldots, g_{\kappa_{2N+1}})$, whose elements are expressed as a linear combination of complex exponentials



\begin{equation}
    g_{\kappa_n}=\sum_{p=1}^Pa_p\lambda_p^{\kappa_n},
\end{equation}
with $n=1,2,\ldots,2N+1$.

Using Kronecker's theorem for complex symmetric matrices, the solution to this approximation problem can be expressed in terms of $\bn{g}$, whose associated Hankel matrix has rank $P$ and minimizes the residual between $\bn{x}$ and~$\bn{g}$. 

For the sake of completeness, let's recall that a Hankel matrix $\bn{H}\!=\!\mathcal{H}(\bn{g})$, generated element-wise from a vector $\bn{g}$, is a symmetric matrix where each ascending skew diagonal from left to right is constant such that $[\bn{H}]_{ij}\!=\!g_{\kappa_{i+j-1}}$ with $1\leq i,j\leq N+1$. On the other hand, Kronecker's theorem \cite{Rochberg1987,Andersson2011} for complex symmetric matrices establishes that $\operatorname{rank}(\bn{H})\!=\!P$ if and only if there exist $\{\lambda_p\}_{p=1}^P$ and $\{a_p\}_{p=1}^P$ in $\mathbb{C}$ such that $g_\kappa\!=\!\sum_{p=1}^P a_p \lambda_p^\kappa$ for equally spaced indices $\kappa=\kappa_{1},\ldots, \kappa_{2N+1}$.

Consequently, the best approximation (in the $l_2$-norm sense) of $\bn{x}$ by a linear combination of $P$ complex exponentials is given by the vector $\bn{g}$ which satisfies $\operatorname{rank}(\mathcal{H}(\bn{g})) = P$ and minimizes the $l_2$-norm of the difference $\bn{x}-\bn{g}$. Inspired by \cite{Andersson2014}, $\bn{g}$ is obtained as the solution of the optimization problem

\begin{equation}\label{eq: optim}
\begin{aligned}
& \underset{\bn{H},\,\bn{g}}{\operatorname{minimize}}\quad {\frac{1}{2}\,\|\bn{x}-\bn{g}\|_{\bn{W}}^2} + \mathcal{R}_P(\bn{H}), \\
& \operatorname{subject\,to}\quad \bn{H}=\mathcal{H}(\bn{g}),
\end{aligned}
\end{equation}
where $\mathcal{R}_P(\bn{H})$ is an indicator function that equals zero if $\operatorname{rank}(\bn{H}) \leq P$, and infinity otherwise, {$
\|\bn{x} - \bn{g}\|_{\bn{W}}^2\! =\! (\bn{x} - \bn{g})^{\mathrm{H}} \bn{W} (\bn{x} - \bn{g})$
 stands for the $\bn{W}$-weighted $l_2$-norm and $\cdot^H$ is the conjugate transpose. The matrix $\bn{W} \!=\! \operatorname{diag}(\bn{w})$ is a diagonal matrix with non-negative entries defined by the weight vector $\bn{w} = (w_{\kappa_1}, \ldots, w_{\kappa_{2N+1}})$. In Section~\ref{Weight}, we propose an iteratively reweighted approach to mitigate noisy data, in contrast to our previous approach introduced in~\cite{leon2024isbi}, which assumes no weights, i.e., $\bn{w} \equiv \bn{1}$.}


\subsection{ADMM-based solution}\label{sec: adm}

The optimization problem~\eqref{eq: optim} can be solved using the alternating direction method of multipliers (ADMM) method~\cite{Boyd2011}, an iterative technique where a solution to a large global problem is obtained by solving smaller sub-problems. To solve \eqref{eq: optim}, the ADMM algorithm transitions from iteration $q$ to iteration $q+1$ as follows 
\begin{align}
    \bn{H}^{q+1}\!&=\!\underset{{\bn{H}}}{\argmin}\!~\mathcal{L}(\bn{H},\bn{g}^q,\bn{\Lambda}^q), \label{eq: admm1} \\ 
    \bn{g}^{q+1}\!&=\!\underset{\bn{g}}{\argmin}\!~ \mathcal{L}(\bn{H}^{q+1},\bn{g},\bn{\Lambda}^q), \label{eq: admm2} \\
    \bn{\Lambda}^{q+1}\!&=\!\bn{\Lambda}^q +{\rho}(\bn{H}^{q+1}-\mathcal{H}(\bn{g}^{q+1})), \label{eq: admm3}
\end{align}
 where $\bn{\Lambda}$ is the Lagrange multiplier matrix, and 
 \vspace{-2mm}
 \begin{equation}\label{eq: lag}
\begin{aligned}
\mathcal{L}(\bn{H},\bn{g},\bn{\Lambda})\!=&\!\mathcal{R}_P(\bn{H})+{\frac{1}{2}\,\|\bn{x}-\bn{g}\|_{\bn{W}}^2}\\
& +\langle \bn{\Lambda},\bn{H}-\mathcal{H}(\bn{g})\rangle_{\text{Re}} +\frac{\rho}{2}||\bn{H}-\mathcal{H}(\bn{g})||_F^2,
\end{aligned}
\vspace{-2mm}
\end{equation} 
is the augmented Lagrangian associated with~\eqref{eq: optim}. In \eqref{eq: lag}, $\rho$ is a penalty coefficient, 
$\langle \bn{A},\bn{B}\rangle_{\text{Re}}\!=\!\Re(\operatorname{tr}(\bn{A}\bn{B}^H))$ denotes the inner product between the matrices $\bn{A}$ and $\bn{B}$, and $||\bn{A}||^2_F={\langle \bn{A},\bn{A}\rangle}_{\text{Re}}$ is the Frobenius norm. 

 



 
    
    
    

         
    
    

   

The ADMM procedure defined in~\eqref{eq: admm1}-\eqref{eq: admm2}-\eqref{eq: admm3} ends after a finite number $Q$ of iterations. 
The rank constraint makes~\eqref{eq: optim} a non-convex problem, meaning that convergence to a unique solution is not guaranteed and may depend on the choice of parameters $\rho$ and $Q$ as well as the initial values $\bn{g}^0$ and $\bn{\Lambda}^0$. Nevertheless, results in~\cite{Andersson2014} show that the Hankel-based spectral method performs substantially better than established high-resolution techniques like ESPRIT~\cite{Esprit}. In this work, we fixed the parameters to $Q\!=\!200$, $\rho\!=\!0.025$, $\bn{g}^{0}\!=\!\bn{x}$ and $\bn{\Lambda}^{0}\!=\!\bn{0}$, following the configuration in~\cite{Andersson2014, leon2024isbi}.

The most computationally intensive operation in each iteration step $q$ is the singular value decomposition (SVD) performed during the update of $\bn{H}$. The time complexity of the SVD operation is $\mathcal{O}(N^3)$, leading to an overall time complexity for the algorithm of $\mathcal{O}(QN^3)$. In comparison, the ESPRIT method requires the computation of only a single SVD, while the AR approach~\cite{Rohrbach2018} does not require any SVD computations.

Due to the regularization term $\frac{\rho}{2}||\bn{H}-\mathcal{H}(\bn{g})||_F^2$ in~\eqref{eq: lag}, it is common that $\bn{\hat{H}}\!\neq \!\mathcal{H}(\bn{\hat{g}})$, where $\bn{\hat{H}}$ and $\bn{\hat{g}}$ are estimates of $\bn{H}$ and $\bn{g}$ resulting from the ADMM algorithm.  This discrepancy suggests that $\bn{\hat{g}}$ or the vector ${\bn{x}}^{\text{approx}}$, obtained by averaging the anti-diagonals terms of $\bn{\hat{H}}$ provides a valid approximation of a sum of exponential functions and can serve as a solution to~\eqref{eq: optim}.
The vector $\bn{x}^{\text{approx}}$ is defined element-wise as
\begin{equation}
    x^{\text{approx}}_{\kappa_n}\!=\!\frac{1}{\mu_{n}}\underset{i+j=n+1}{\sum}[\bn{\hat{H}}]_{ij},
\end{equation}
where $\mu_{n}$ is the number of times that ${\hat{g}}_{\kappa_n}$ appears in the matrix $\bn{\hat{H}}$.
Finally, the latter $\bn{x}^{\text{approx}}$ is considered a more appropriate solution because $\bn{\hat{g}}$ is computed to best approximate the noisy data.

\subsection{Spectral parameter estimation}

By construction, the Hankel matrix solution $\bn{H}\!=\! \mathcal{H}(\bn{x}^{\text{approx}})$ is a complex symmetric matrix which satisfies $\bn{H}\!=\!\bn{H}^T$ and $\operatorname{rank}(\bn{H})\!=\!P$. For matrices with these properties, one can choose the matrices $\bn{U}$ and $\bn{V}$ appearing in the SVD of $\bn{H}$, i.e., $\bn{H}\!=\!\bn{U}\bn{\Sigma}\bn{V}^H$ such that $\bn{U}\!=\!\overline{\bn{V}}$, where the bar indicates complex conjugation. Consequently, 
\begin{align}
\bn{H}\!=\!\sum_{p=1}^{P}{s}_p\bn{u}_p\bn{u}_p^T, \, s_p\in \mathbb{R}^+,\, \bn{u}_p\in \mathbb{C}^{2N+1},
\end{align}
 where ${s}_1,{s}_2,\ldots ,{s}_{P}>0$ are the 
 eigenvalues of $\bn{H}$. Moreover, the con-eigenvectors $\bn{u}_p$ are orthogonal, satisfy the relation $\bn{H}\overline{\bn{u}}_p\!=\!s_p\bn{u}_p$ and can be expressed as the sum of the same $P$ exponentials of $\bn{x}^{\text{approx}}$.
 
Hence, we can decompose the matrix $\bn{U}\!=\!(\bn{u}_1 , \ldots , \bn{u}_P )$ as $\bn{U}\!=\! \bn{E}\bn{G}$, where $\bn{G}$ is some invertible $P \times P$ matrix and  $\bn{E}$ is a $2N + 1 \times P$ Vandermonde-like matrix defined by $[\bn{E}]_{np}=\lambda_p^{\kappa_{n}}$, for {$p\!=\!1,\ldots,P$ and $n\!=\!1,\ldots,2N+1$}, i.e., 
\begin{equation}\label{matrix E}
\bn{E}=\left(\begin{matrix}
 \lambda_1^{\kappa_1}& \ldots &\lambda_P^{\kappa_1}\\
 \vdots& \ddots &\vdots\\
 \lambda_1^{\kappa_{2N+1}}& \ldots & \lambda_P^{\kappa_{2N+1}}\\
\end{matrix}\right).
\end{equation}

Let $\bn{U}^f$ (resp. $\bn{U}^l$) denote the matrix $\bn{U}$ 
whose first row (resp. last row) has been dropped. Since $\bn{U}^f \!=\! \bn{E}^f \bn{G}$, $\bn{U}^l \!=\! \bn{E}^l \bn{G}$ and given the structure of $\bn{E}$, it is possible to obtain $\bn{E}^f\!=\!\operatorname{diag}(\lambda_1 , \ldots , \lambda_P )\bn{E}^l$. 
Consequently, 
$$(\bn{U}^l)^\dagger\bn{U}^f=\bn{G}^{-1}\operatorname{diag}(\lambda_1 , \ldots , \lambda_P )\bn{G}$$
where $\cdot^\dagger$ denotes the pseudo-inverse operator.
Therefore, $\{\lambda_p\}_{p=1}^P$ can be obtained by computing the eigenvalues of $(\bn{U}^l)^\dagger\bn{U}^f$.

Finally, the amplitudes $\{a_p\}_{p=1}^P$ are estimated using 
the LS estimator
\begin{equation}
\bn{\hat{a}}=(\bn{E}^T\bn{E})^{-1}(\bn{E}\bn{x}),
\end{equation}
where $\bn{\hat{a}}=(\hat{a}_1,\hat{a}_2,\ldots,\hat{a}_P)$.

\subsection{Iteratively reweighted approach}\label{Weight}

We propose to use a weight function based on redescending M-estimators to reduce the impact of noisy data points on estimates. Specifically, we adopt the Tukey bisquare weight function \cite{Beaton1974,Nora2018}, which robustly rejects outliers, defined as
\begin{equation}\label{w_k}
    w_\kappa=\begin{cases}
        (1-(e_\kappa/(r\delta))^2)^2 \text{ if $|e_\kappa|\leq c\delta$,}        \\
        0  \text{ otherwise,}
        \end{cases}
\end{equation}
where $e_\kappa$ is the residual error at the data point $\kappa$ and $c > 0$ is a constant fixed a priori. The scale parameter  $\delta > 0$  represents the standard deviation of the residual errors for the inlier estimates, and is generally estimated using a robust estimator such as the median absolute deviation (MAD) \cite{Meer1991}
$$\hat{\delta}_{\text{MAD}}=\delta_0\underset{n=1,\ldots,2N+1}{\operatorname{median}}{(|e_{\kappa_n}-\operatorname{median}(\bn{e})|)}$$
with $\bn{e} = (e_{\kappa_1} ,\ldots, e_{\kappa_{2N+1}} )$ and $\delta_0= 1.4826$ is a scaling constant for Gaussian errors \cite{Odobez1995}.

The proposed iteratively reweighting strategy starts with initial weights set to 1. After performing spectral estimation, the residuals from the spectral reconstruction are calculated, and the weights are updated according to~\eqref{w_k}. This iterative process gradually down-weights data points with high residuals, thereby reducing their influence on the estimation and improving robustness in the presence of noise.

Our experiments indicate that repeating this process three times provides a good balance between computational efficiency and estimation accuracy.


\section{Cramér-Rao Bounds} \label{sec:CRB}

The Cramér-Rao bound (CRB) \cite{VanTrees2001} is a theoretical lower bound on the variance of any unbiased estimator for a given parametric model. The CRB provides a valuable benchmark for assessing the performance of estimation algorithms, as it defines the best achievable performance given the statistical characteristics of the data and the noise. In addition, deriving this bound offers insight into the estimation problem by identifying which parameters are more difficult to estimate and under what conditions. It also helps corroborate the underlying physical hypothesis and establish experimental conditions that enable accurate parameter estimation.  

In this section, we derive the CRB for the estimation of the acoustic parameters within the framework established in Section~\ref{sec: model}. First, we present the CRB for the spectral parameters of the model defined in~\eqref{general process}. Then, leveraging the functional invariance property, we derive the CRB for the acoustic parameters of interest. This final derivation represents a key contribution of this work.


\subsection{Bounds for spectral parameter estimation}

The CRB for the real-valued parameter vector $$\bn{\theta}\!=\!(A_1,\ldots,A_P,b_1,\ldots,b_P,\gamma_1,\ldots,\gamma_P,\nu_1,\ldots,\nu_P)$$ in spectral model~\eqref{general process} is defined by the diagonal elements of the inverse of the Fisher Information Matrix (FIM) $\bn{F}(\bn{\theta})$ \cite{VanTrees2001}, i.e., 
\begin{equation}
\operatorname{CRB}(\bn{\theta}_i) = [\bn{F}^{-1}(\bn{\theta})]_{ii}.
\end{equation}




We assume that the error $\epsilon_\kappa$ in~\eqref{general process} is zero mean, uncorrelated, and follows a complex Gaussian distribution, with independent real and imaginary components, each having variance $\sigma^2/2$. Under these conditions, the FIM can be computed using the Slepian-Bang’s formula \cite{stoica2005spectral}, i.e.,
\begin{equation}\label{fisher}
\bn{F}(\bn{\theta})=\frac{2}{\sigma^2}\Re\left(\left(\dfrac{\partial \bn{g}}{\partial \bn{\theta}}\right)^H\left(\dfrac{\partial \bn{g}}{\partial \bn{\theta}}\right)\right).
\end{equation}
Following the presentations in \cite{Gudmundson2012} and \cite{YingXianYao1995}, we use the derivatives $$
\dfrac{\partial g_\kappa}{\partial {A_p}}\!=\!\exp{(\mathrm{i}b_p)}\lambda_p^\kappa, \, \dfrac{\partial g_\kappa}{\partial {\beta_p}}\!=\!\mathrm{i}a_p\lambda_p^\kappa, \, \dfrac{\partial g_\kappa}{\partial {\gamma_p}}\!=\!2\pi\kappa_p\lambda_p^\kappa \text{\, and \,}  \dfrac{\partial g_\kappa}{\partial {\nu_p}}\!=\!\mathrm{i}2\pi\kappa a_p\lambda_p^\kappa$$ to form the matrix 
\begin{equation}\label{patial g}
\bn{Q}\!=[\bn{E B} \quad \mathrm{i}\bn{E B } \quad  2\pi\bn{\tilde{E} B } \quad  \mathrm{i}2\pi\bn{\tilde{E} B }],
\end{equation}
where $\bn{E}$ is defined in~\eqref{matrix E},
\begin{equation*}
\bn{\tilde{E}}=\left(\begin{matrix}
 \kappa_1\lambda_1^{\kappa_1}& \ldots &\kappa_1\lambda_P^{\kappa_1}\\
 \vdots& \ddots &\vdots\\
 \kappa_{2N+1}\lambda_1^{\kappa_{2N+1}}& \ldots & \kappa_{2N+1}\lambda_P^{\kappa_{2N+1}}\\
\end{matrix}\right),
\end{equation*}
and $\bn{B}\!=\!\operatorname{diag}(\exp{(\mathrm{i}b_1)},\ldots,\exp{(\mathrm{i}b_P)})$. Therefore, we can express the inverse FIM as
 \begin{align}\label{fisher2}
 \bn{F}^{-1}(\bn{\theta})=\frac{\sigma^2}{2}\bn{P}^{-1}\left(\Re\left( \bn{Q}^H \bn{Q}\right)\right)^{-1}\bn{P}^{-1},
\end{align}
where $\bn{P}=\operatorname{diag}[ \bn{I}\, \bn{A}\, \bn{A}\, \bn{A}]$ with $\bn{A}\!=\!\operatorname{diag} (a_1,\ldots,a_p)$.



\subsection{Bounds for acoustic parameter estimation}


As observed in \eqref{eq: c and Z} and \eqref{eq d and alpha}, the acoustic parameters are determined through transformations of the spectral parameters $\bn{\theta}$. We denote these transformations as $\bn{\phi}\!=\!(\phi_1,\phi_2,\phi_3,\phi_4)$, where $\phi_1(\bn{\theta})\!=\!c$, $\phi_2(\bn{\theta})\!=\!d$, $\phi_3(\theta)\!=\!\alpha$ , $\phi_4(\bn{\theta})\!=\!Z$. Using the functional invariance property of the CRB \cite{VanTrees2001}, we obtain the CRB for the acoustic parameters as the diagonal elements of the inverse of the transformed FIM, denoted by $\bn{F}^{-1}_{\bn{\phi}}(\bn{\theta})$ and computed as 
\begin{equation}
\bn{F}^{-1}_{\bn{\phi}}(\bn{\theta})=\nabla_\theta[\bn{\phi}^T(\bn{\theta})]^T \bn{F}^{-1}(\bn{\theta})\nabla_\theta[\bn{\phi}^T(\bn{\theta})],
\end{equation}
where $\nabla_\theta[\bn{\phi}^T(\bn{\theta})]=\left(\dfrac{\partial\bn{\phi}_1(\bn{\theta})}{\partial\bn{\theta}},\dfrac{\partial\bn{\phi}_2(\bn{\theta})}{\partial\bn{\theta}},\dfrac{\partial\bn{\phi}_3(\bn{\theta})}{\partial\bn{\theta}},\dfrac{\partial\bn{\phi}_4(\bn{\theta})}{\partial\bn{\theta}}\right)$ is the Jacobian matrix of $\bn{\phi}^T(\bn{\theta})$ with respect to $\bn{\theta}$, and
\begin{align*}    
    \dfrac{\partial\bn{\phi}_1(\bn{\theta})}{\partial\bn{\theta}}&=(\bn{0}_{3P},~c_w\nu_2/(\nu_1-\nu_2)^2,c_w\nu_1/(\nu_1-\nu_2)^2,~\bn{0}_{P-2}),\\
    \dfrac{\partial\bn{\phi}_2(\bn{\theta})}{\partial\bn{\theta}}&=(\bn{0}_{3P},c_w/(2f_s),~\bn{0}_{P-1}), \\
    \dfrac{\partial\bn{\phi}_3(\bn{\theta})}{\partial\bn{\theta}}\!&=\!(\bn{0}_{2P+1},1/(c_w\nu_1),~\bn{0}_{P-2},\gamma_2/(c_w \nu_1^2),~\bn{0}_{P-1}), \\
   \dfrac{\partial\bn{\phi}_4(\bn{\theta})}{\partial\bn{\theta}}&=(2Z_w/(R_{wg}(1-a_1/R_{wg}))^2,\bn{0}_{4P-1}),
\end{align*}
where $\bn{0}_{K}$ represents a vector of zeros of size $K$.


Finding closed-form expressions for the CRB becomes challenging due to the dense structure of the matrix $\bn{Q}$. Ultimately, we express the CRBs for the acoustic parameters as
\begin{equation}\label{crb_acus}
\begin{aligned}
    \operatorname{CRB}_c&=[\bn{F}^{-1}_{\bn{\phi}}(\bn{\theta})]_{11}, \quad \operatorname{CRB}_d=[\bn{F}^{-1}_{\bn{\phi}}(\bn{\theta})]_{22}, \\
    \operatorname{CRB}_\alpha&=[\bn{F}^{-1}_{\bn{\phi}}(\bn{\theta})]_{33}, \quad \operatorname{CRB}_Z=[\bn{F}^{-1}_{\bn{\phi}}(\bn{\theta})]_{44}.
\end{aligned}
\end{equation}

The only clear dependency that can be extracted from these formulas is that on the noise level. Specifically, all acoustic CRBs in~\eqref{crb_acus} are scaled by a factor of $\sigma^2/2$ coming from~\eqref{fisher2}, indicating that the CRBs increase proportionally to the noise variance. Furthermore, the acoustic CRBs explicitly depend on the spectral values and implicitly on the acoustic parameters themselves. This latter dependency is examined in the results section through simulations.


\section{Material and Methods}\label{sec:material}

QAM systems are capable of operating at frequencies of 250~MHz~\cite{Rohrbach2015retine} and 500~MHz~\cite{Rohrbach500MHZ}. Although increasing the transducer frequency improves spatial resolution, it also introduces heightened sensitivity to environmental factors such as vibrations and temperature fluctuations, increasing the challenges in the estimation of acoustic parameters.

{The Hankel-based methods introduced in Section~\ref{sec:HK}, referred to as RHK for the proposed reweighted approach and HK for the non-weighted version~\cite{leon2024isbi} (which corresponds to the initialization of RHK), along with the standard autoregressive algorithm~\cite{Rohrbach2018}, are applicable across different transducer frequencies. In this study, we focus on real and simulated QAM data acquired using a 500~MHz transducer.}


QAM estimates are considered outliers or unreliable values if they fall outside physically admissible ranges, specifically $c \!< \!1500$ m/s, $c\! > \!2200$ m/s, $Z \!< \!1.48$ MRayl, or $Z\! >\! 2.2$ MRayl~\cite{Rohrbach2018}. Therefore, beyond assessing estimator accuracy {using the root mean square error (RMSE)}, the outlier count serves as an additional metric to evaluate the efficacy of each method. {From this point onward, outliers are excluded from RMSE calculations to ensure a fair comparison across methods.}

\subsection{Simulation setting}

We used a measured reference signal $h_0$ from our 500 MHz QAM system, to simulate QAM RF signals applying the inverse Fourier transform to the model in~\eqref{fourier}, for a given set of acoustic parameters, $P\! =\! 2$, $M\! =\! 300$, and $f_s\! =\! 10\, \text{GHz}$. The resulting signals were then perturbed by independent and identically distributed (i.i.d.) zero mean Gaussian noise with variance $\sigma_x^2$, given by  
\begin{equation}\label{sigma}
    \sigma_x\!=\!10^{(20 \log_{10}((\xi(h_0)-\text{SNR})/20))},
\end{equation}
where \(\xi(h_0)\) is the maximum of the absolute value of the envelope of $h_0$, estimated using its Hilbert transform-based analytic signal~\cite{Rohrbach2018}. 
This noise formulation ensures that the simulated signals maintain the desired signal-to-noise ratio (SNR) in the time domain, representing the difference in decibels (dB) between the amplitude of the reference signal and the added noise. To assess estimation performance, we generated 200 realizations of the random Gaussian noise for each parameter configuration.  


In each simulation, the acoustic parameters were set to experimentally relevant values for QAM applications, chosen based on preliminary tests to determine the optimal range between easily separable scenarios (e.g., large SNR, $d$ and $|Z-Z_w|$, and small $\alpha$) with more challenging cases (e.g., small SNR, $d$, and $|Z-Z_w|$, and large $\alpha$)~\cite{Rohrbach2018}. We varied the parameter under investigation while keeping the remaining parameters fixed as follows: 
SNR $\!=\! 50$ dB, $d \!=\! 4$ $\mu$m, $Z \!=\! 1.63$ MRayl, $c \!=\! 1600$ m/s,  $\alpha \!=\! 10$ dB/MHz/cm. We evaluated the effect of decreasing the SNR, reducing signal separation (i.e., thinner samples), lowering the amplitude of the first signal (i.e., acoustic impedance contrast $|Z-Z_w|$), and increasing sample attenuation. Table~\ref{tab: parameter values} summarizes the ranges of all parameter values used in our simulations.

\begin{table}[t]
\caption{Simulated parameter variations}\label{tab: parameter values}
\vspace{-5mm}
    \begin{center}
    \begin{tabular}{|c|c|c|c|}
    \hline
        Parameter & Lower value & Step size & Upper value            \\ \hline\hline
         SNR [dB] & 20         & 5         & 100                    \\ \hline
         $Z$  [MRayl]                & 1.51       & 0.01      & 1.63 \\ \hline
         $\alpha$ [dB/MHz/cm]        & 8          & 1         & 20   \\ \hline
         $d$ [$\mu$m]                & 1          & 0.5         & 8    \\ \hline
         bandwidth [dB]          & -4    & -2         & -20   \\ \hline
    \end{tabular}
\end{center}
\vspace{-3mm}
\end{table}

\subsection{CRB approximation}

In QAM simulations, i.i.d. zero-mean Gaussian noise with variance $\sigma_x^2$ is typically added in the time domain as described before. However, due to the division by Fourier coefficients during the transformation to the frequency domain, the noise no longer maintains its identical distribution. This discrepancy violates the assumptions made in Section~\ref{sec:CRB} for deriving the CRBs. To address this limitation, we consider an adjusted bound that approximates the power of the non-identically distributed noise with that of an identically distributed noise. Specifically, for each simulated signal, we extract the noise present in the truncated normalized spectrum and compute its sample variance. Then, the value of $\sigma^2$ used in Section~\ref{sec:CRB} is approximated by the average of these sample variances, calculated over a large number (e.g., 5,000) of simulated signals. Our experiments demonstrated that this practical adjustment still provides a reliable lower bound for studying the behavior of the estimators.

\subsection{\emph{Ex vivo} experiments}

{Experimental data were acquired using the 500~MHz QAM system described in~\cite{Rohrbach500MHZ}. Two biological specimens were analyzed: a 6~$\mu$m-thick section of human lymph node and a 6~$\mu$m-thick section of pig cornea. Data acquisition was performed at a sampling frequency of 10~GHz. At each spatial location, a 300-point RF signal was recorded using a lateral step size of 1~$\mu$m in both directions.

For analysis, regions of interest (ROIs) measuring $60\!\times\!90~\mu$m (lymph node) and $38\!\times\!99~\mu$m (pig cornea) were selected.}

\subsection{Preprocessing} \label{prepo}

Before applying any estimation method, a Cadzow filter \cite{Cadzow1988} is used to denoise the normalized spectrum $\bn{x}$. This process aims to reduce the rank of the Hankel matrix $H(\bn{x})$ to $P$. Specifically, the SVD of $H(\bn{x})$ is computed and $H(\bn{x})$ is then reconstructed by retaining only the $P$ largest singular values. Then, a denoised version of $\bn{x}$ is obtained by averaging all the anti-diagonals of the rank-$P$ approximation of $H(\bn{x})$. This procedure is repeated five times \cite{Rohrbach2018}.

In \cite{leon2024isbi}, we investigated whether this preprocessing step is essential for the non-weighted Hankel method. While its use was important in simulations, its impact on real data was minimal. Nonetheless, we continue to apply the Cadzow filter to ensure fairness when comparing performance with the AR-based method, which requires this denoising step.

The model order $P$ can either be fixed (e.g., $P \geq 2$) or dynamically determined during the first iteration of the Cadzow filter, based on the number of singular values that contribute more than 10$\%$ to the total signal energy. This dynamic approach provides a more tailored representation of the underlying signal and is used for real data. For simulations, we set $P = 2$ because the CRB calculations depend on this value and the associated (known) spectral parameters.


\begin{figure}[tb]
    \centering
    \begin{subfigure}[b]{0.45\textwidth}
        \caption{}
        \includegraphics[width=\textwidth, trim=20 20 20 20,clip]{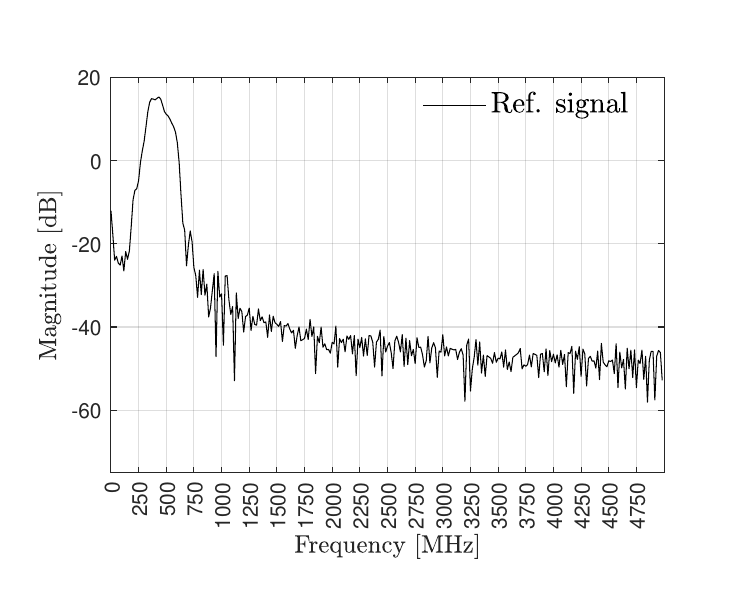}  
        \label{fig:sub1}
    \end{subfigure}
    \hfill
    \begin{subfigure}[b]{0.45\textwidth}
        \caption{}
        \includegraphics[width=\textwidth, trim=20 20 20 20,clip]{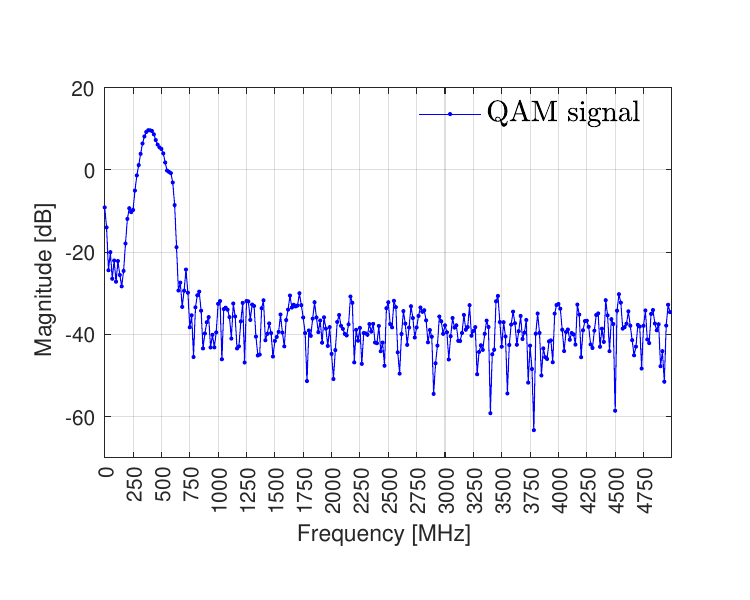}
        \label{fig:sub2}
    \end{subfigure}
    
   \vspace{-0.5cm}  

    \begin{subfigure}[b]{0.45\textwidth}
        \caption{}
        \includegraphics[width=\textwidth, trim=20 20 20 20,clip]{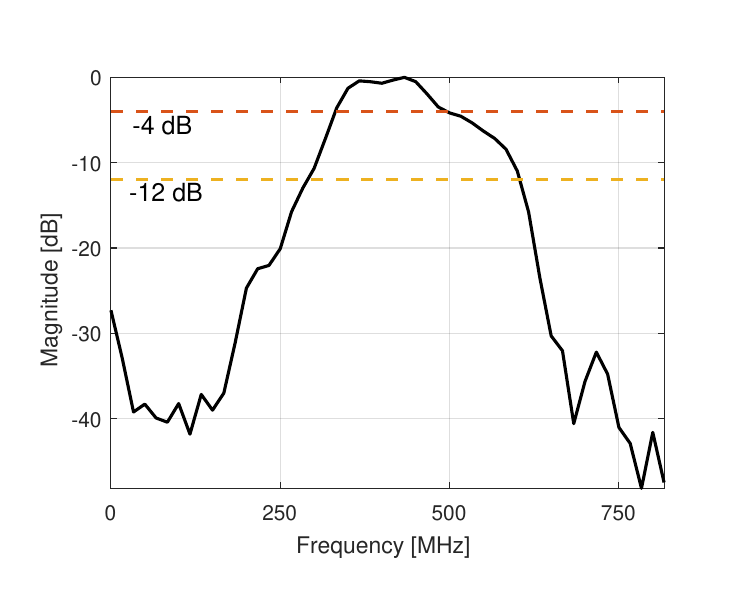}
        \label{fig:sub3}
    \end{subfigure}
    \hfill
    \begin{subfigure}[b]{0.45\textwidth}
        \caption{}
        \includegraphics[width=\textwidth, trim=0 20 20 20,clip]{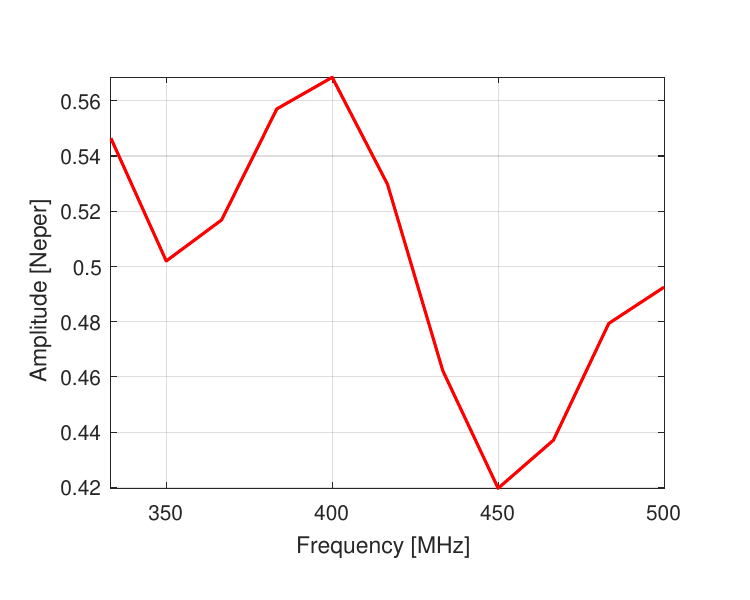}
        \label{fig:sub4}
    \end{subfigure}
    
    \caption{{(a) Half-spectrum of the reference signal (substrate echo). 
(b) Half-spectrum of the QAM signal acquired from a sample region. 
(c) Normalized magnitude spectrum with -4 dB and -12 dB thresholds used for bandwidth estimation. 
(d) Ratio of QAM to reference spectra within the -4 dB bandwidth window.}
}
    \label{fig:band}
\end{figure}

The choice of bandwidth determines the range of frequencies where division in \eqref{fourier} is well-defined and affects the trade-off between the number of data points and the noise level in the truncated spectrum. The usable bandwidth typically includes frequencies within -6 to -12 dB of the center (or peak) frequency, depending on the sensitivity of the transducer and system technology, as obtained from $\mathcal{F}_{h_0}$ \cite{Hoerig2023}. {To illustrate the bandwidth selection, Fig.~\ref{fig:band} shows example spectra used in our processing. The top plots correspond to the echo from the reference (no sample) and from a region with a sample. In the bottom row, we show how the bandwidth is defined using the -4 dB and -12 dB thresholds, and how the ratio is computed inside the selected frequency range.} The impact of this choice is analyzed in both simulations and real measurements.

\section{Results}\label{sec:results}

\subsection{Simulations}

\begin{figure}[tb]
    \centering
    \begin{subfigure}[t]{0.6\textwidth}
        \caption{} \label{fig:estimation_rmse}
        \centering
        \includegraphics[width=0.46\textwidth, trim=13 8 29 10,clip]{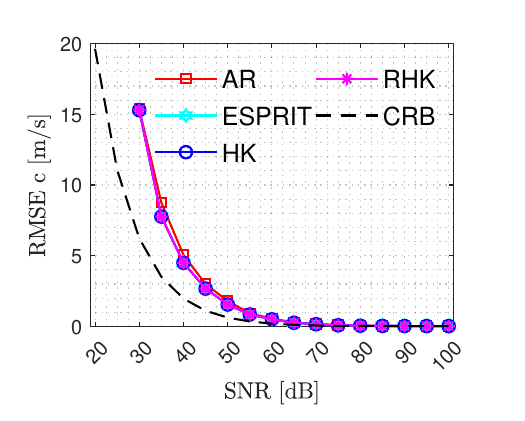}
        \includegraphics[width=0.46\textwidth, trim=13 8 29 10,clip]{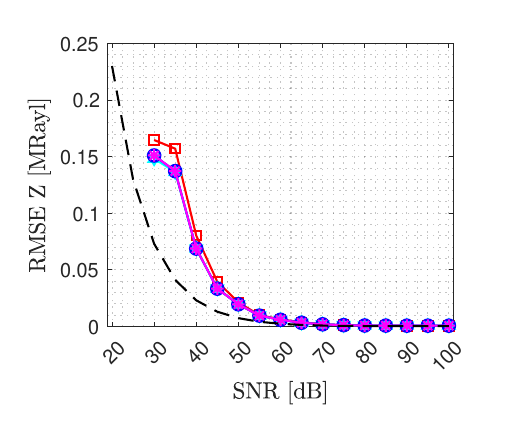}\\
        \includegraphics[width=0.46\textwidth, trim=13 8 29 10,clip]{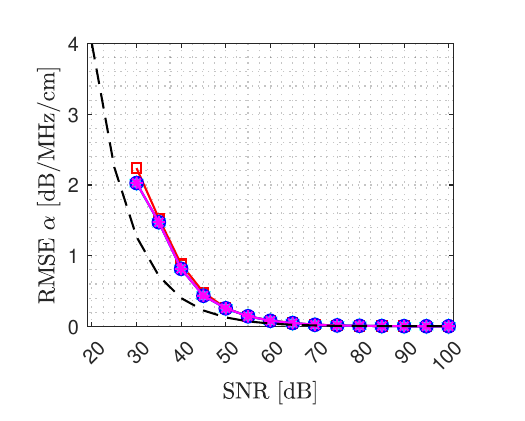} 
        \includegraphics[width=0.46\textwidth, trim=13 8 29 10,clip]{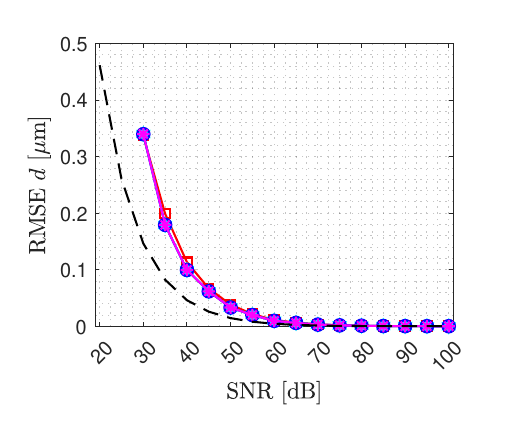}
       
    \end{subfigure}
    \hfill
    \begin{subfigure}[t]{0.39\textwidth}
         \caption{}\label{fig:estimation_outliers}
        \centering
        \includegraphics[width=\textwidth, trim=0 10 0 50,clip]{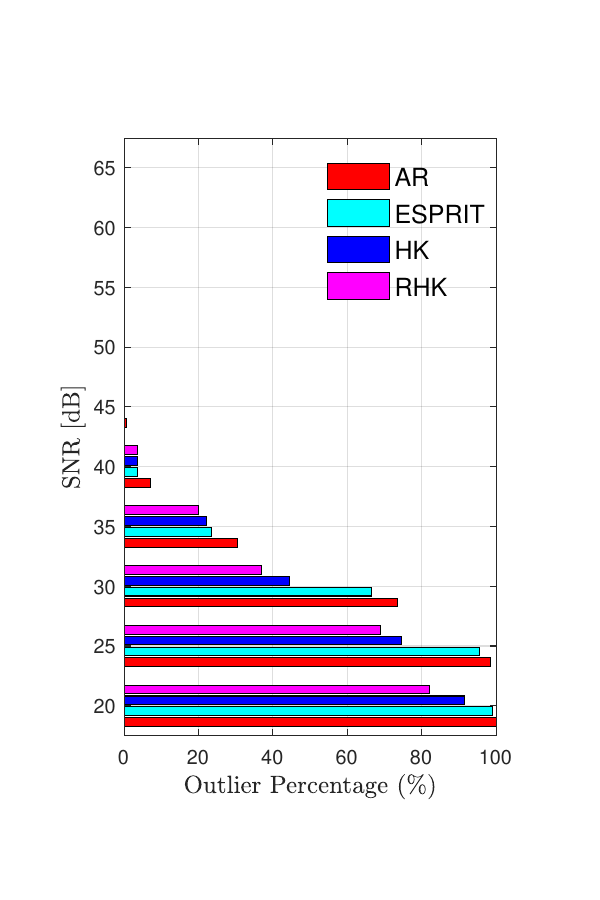}
    \end{subfigure}
    \caption{{Estimation performance versus SNR using simulated 500 MHz QAM data. (a) RMSE for each estimated parameter ($c$, $Z$, $\alpha$, and $d$), after removal of outliers. The proposed RHK is shown in magenta, HK in blue, the AR in red,  ESPRIT in cyan, and the square root of the CRB in black dashed line. (b) Percentage of outliers detected for each method  versus SNR.} }
    \label{fig:est_and_out}
    \vspace{-3mm}
\end{figure}

{We first evaluate the performance of the proposed methods using simulated data across varying SNR levels. Fig.~\ref{fig:est_and_out} presents results for four estimators: proposed weighted Hankel (RHK, magenta), {non-weighted Hankel~\cite{leon2024isbi} (HK, blue)}, {autoregressive~\cite{Rohrbach2018} (AR, red)}, and {ESPRIT~\cite{Esprit} (cyan)}.}

{Fig.~\ref{fig:est_and_out} (a) shows the RMSE for each estimated parameter ($c$, $Z$, $\alpha$, and $d$) as a function of SNR. As expected, RMSE and the corresponding CRB both decrease with increasing SNR for all parameters. At high SNR values (above 60 dB), noise has little impact, allowing all methods to approach the theoretical limit given by the CRB. However, at lower SNRs, the discrepancy between RMSE and CRB becomes more pronounced, with slightly lower RMSE for Hankel methods and ESPRIT.}

{The percentage of outliers obtained for each method is reported in Fig.~\ref{fig:est_and_out} (b). At low SNRs (e.g., 20-30 dB), all methods experience increased outlier rates, yet the Hankel methods demonstrate greater robustness. Specifically, at 20, 25, and 30 dB, RHK yields outlier rates of 84.5\%, 66.5\%, and 47\%, respectively, while HK achieves 88.5\%, 76.5\%, and 49.5\%. In contrast, ESPRIT and AR produce substantially higher outlier rates (ESPRIT: 99.0\%, 95.5\%, 66.5\%; AR: 99.5\%, 98.0\%, 76.5\%).}

{At 35-40 dB, RHK, HK and ESPRIT, have similar outlier rates close to 20\% and 3.5\%, respectively, while AR reported 27.0\% and 6.0\%.}

{Overall, ESPRIT and AR are more sensitive to noise and tend to produce a higher number of outliers under low SNR conditions. In contrast, Hankel approaches, particularly RHK, maintain better robustness, highlighting the advantage of the reweighting strategy in suppressing noise-induced estimation failures. Based on these findings, the remainder of our analysis focuses on comparing AR, as the established baseline for QAM.}

\begin{figure}[!htb]
    \centering
    \includegraphics[width=0.24\textwidth, trim=13 8 29 10,clip]{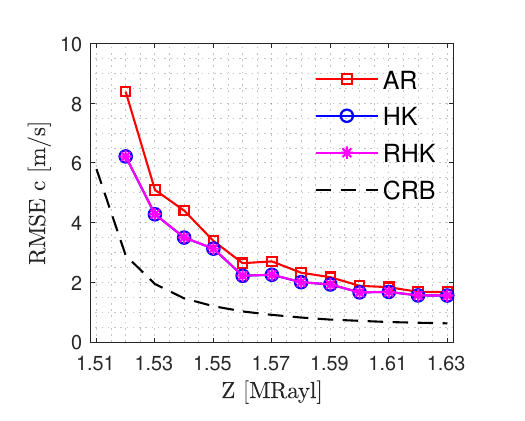}
    \includegraphics[width=0.24\textwidth, trim=13 8 29 10,clip]{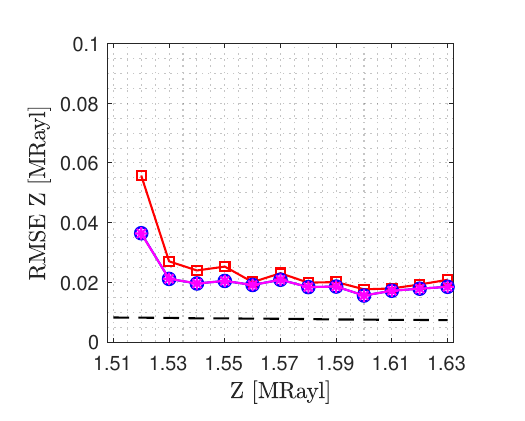}
    \includegraphics[width=0.24\textwidth, trim=13 8 29 10,clip]{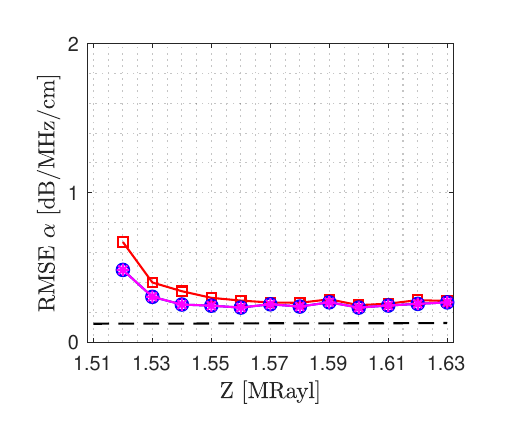} 
    \includegraphics[width=0.24\textwidth, trim=13 8 29 10,clip]{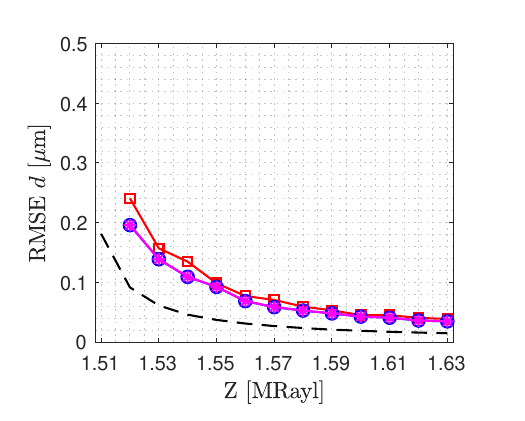}\\
    \includegraphics[width=0.24\textwidth, trim=13 8 29 10,clip]{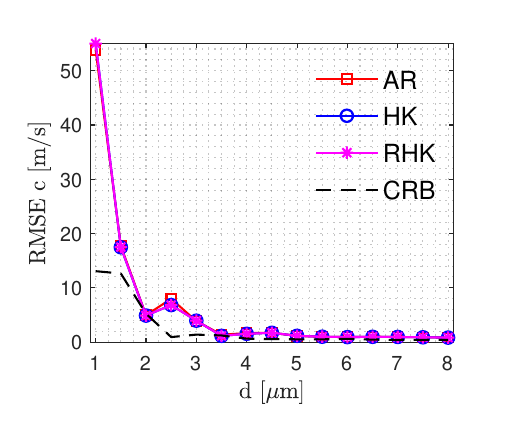}
    \includegraphics[width=0.24\textwidth, trim=13 8 29 10,clip]{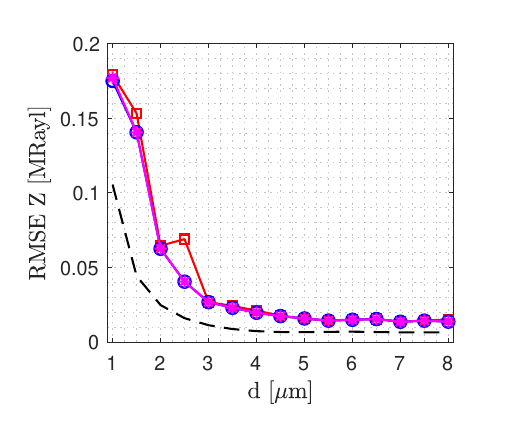}
    \includegraphics[width=0.24\textwidth, trim=13 8 29 10,clip]{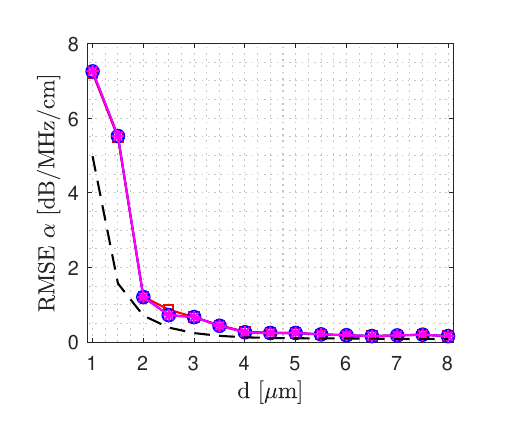} 
    \includegraphics[width=0.24\textwidth, trim=13 8 29 10,clip]{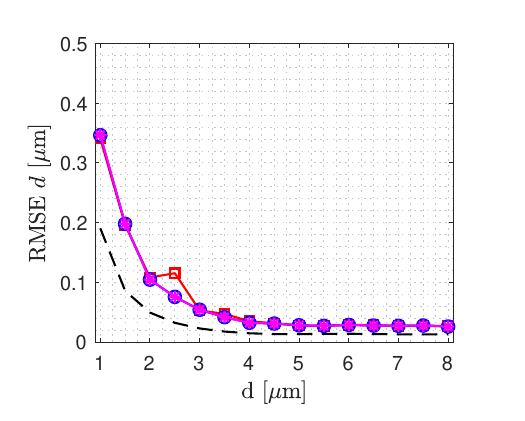}\\
    \includegraphics[width=0.24\textwidth, trim=13 8 29 10,clip]{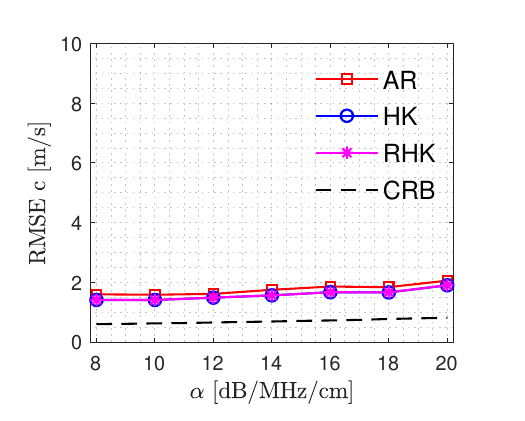}
    \includegraphics[width=0.24\textwidth, trim=13 8 29 10,clip]{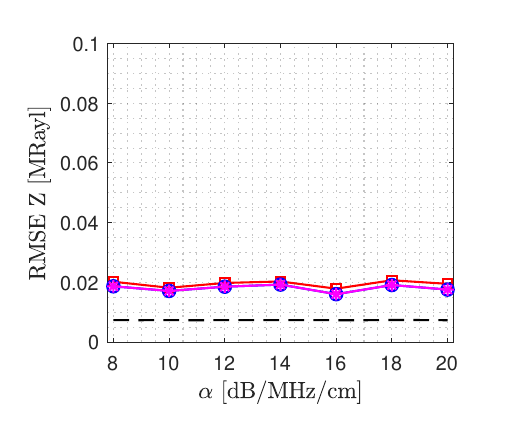}
    \includegraphics[width=0.24\textwidth, trim=13 8 29 10,clip]{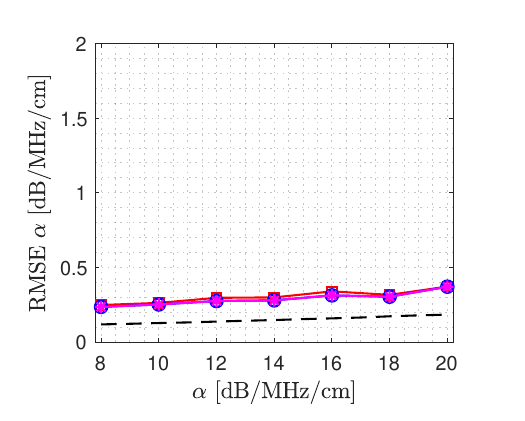} 
    \includegraphics[width=0.24\textwidth, trim=13 8 29 10,clip]{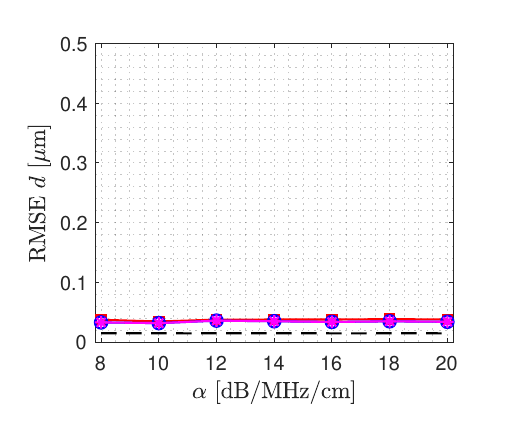}\\
    \includegraphics[width=0.24\textwidth, trim=13 8 29 10,clip]{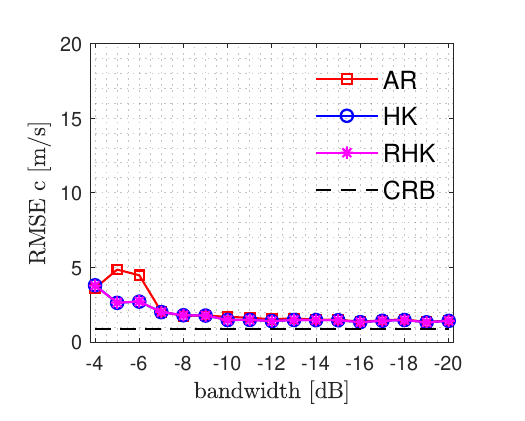}
    \includegraphics[width=0.24\textwidth, trim=13 8 29 10,clip]{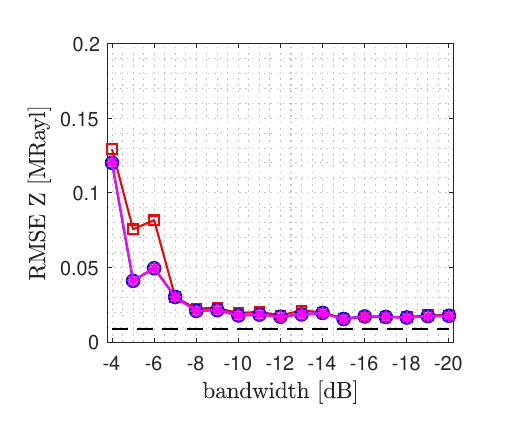}
    \includegraphics[width=0.24\textwidth, trim=13 8 29 10,clip]{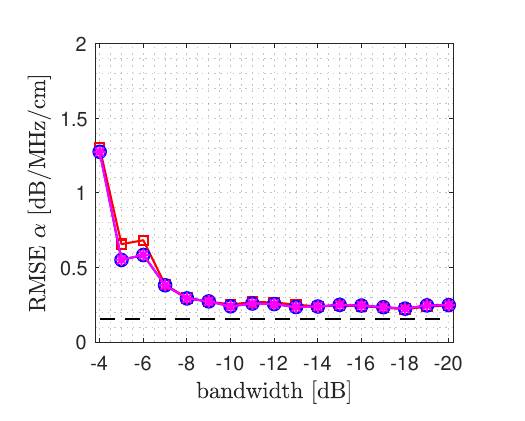} 
    \includegraphics[width=0.24\textwidth, trim=13 8 29 10,clip]{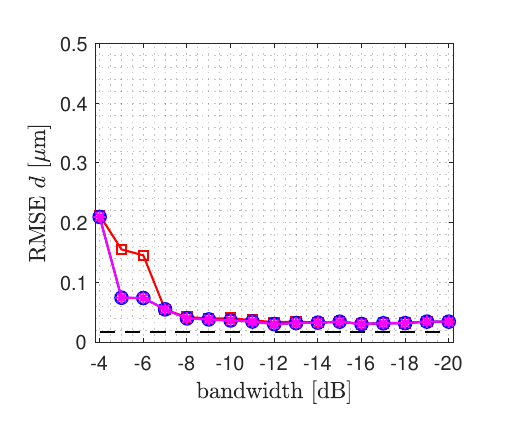}\\
    \caption{Simulations results. Columns represent RMSE results (calculated after outlier removal) for $c$, $Z$, $\alpha$ and $d$. Rows show  different parameter-set variations for $Z$, $\alpha$, $d$ and bandwidth. RHK is in magenta, HK is in blue, AR is in red, and the square root of the CRB is in black dashed line.}\label{fig: estimation}
\vspace{-3mm}
\end{figure}

\begin{figure}[!htb]
    \centering
    \includegraphics[width=0.32\textwidth, trim=0 20 20 20,clip]{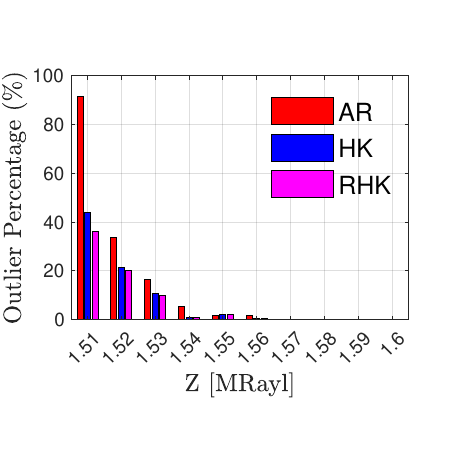}
    \includegraphics[width=0.32\textwidth, trim=0 20 20 20,clip]{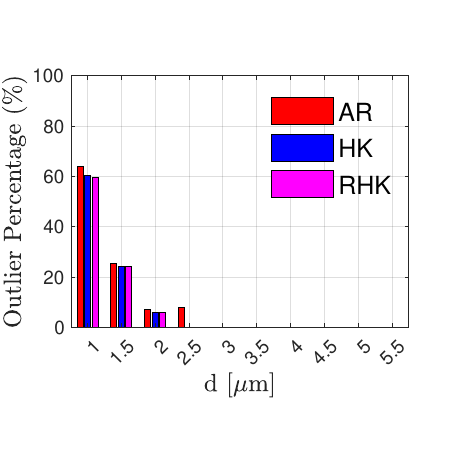}
    \includegraphics[width=0.32\textwidth, trim=0 20 20 20,clip]{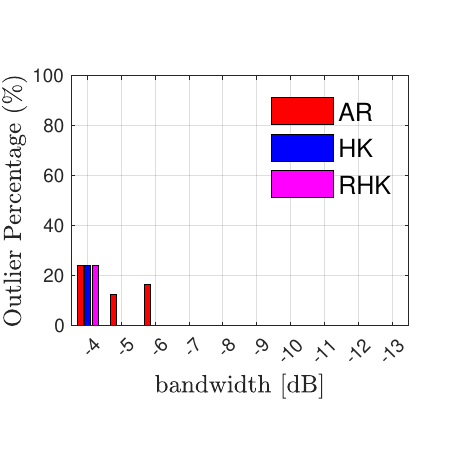}
    \caption{Percentage of outliers detected for $Z$, $d$ and bandwidth variation. RHK is represented in magenta, HK in blue and AR in red.} \label{fig: outliers} 
    \vspace{-3mm}
\end{figure}

{Additional results at a fixed SNR of 50 dB are presented in Fig.~\ref{fig: estimation} and Fig.~\ref{fig: outliers}, which examine RMSE behavior and outliers count as a function of tissue acoustic impedance $Z$, thickness $d$, and attenuation $\alpha$, as well as bandwidth selection.}

Fig.~\ref{fig: estimation} (first row) provides evidence that as $Z$ decreases and approaches the impedance of water ($Z_w \!=\! 1.5$ MRayl), estimation becomes more challenging. This results in large RMSE values and a large gap relative to the CRB for all methods and parameters, which is expected because the amplitude of the water-tissue interface is proportional to $|Z-Z_w|$. However, this trend is not reflected in CRB$_Z$ and CRB$_\alpha$ values, which remain almost constant.
Moreover, Hankel approaches achieve lower RMSE values and a reduced number of outliers compared to AR. In Fig.~\ref{fig: outliers} (left plot), AR shows significantly larger outlier rates of 91$\%$, 33$\%$ and 16$\%$ at $Z \!=\! 1.51$ MRayl, $Z \!=\! 1.52$ MRayl, and $Z\! =\! 1.53$ MRayl, underscoring the improved robustness of Hankel algorithms in these scenarios. At the same $Z$ values, RHK reported outlier rates of 36$\%$, 20$\%$ and 10$\%$, while HK presented increased outlier rates of 44$\%$, 21$\%$ and 10.5$\%$.

In Fig.~\ref{fig: estimation} (second row), RMSE and CRB values decrease as $d$ increases. For all estimators, the RMSE approaches the CRB for large $d$. This confirms that thicker samples yield more accurate parameter estimates, while thinner samples (small $d$) present greater challenges for reliable estimation. Overall, all methods achieve similar RMSE values, except at $d\!=\!2.5$~$\mu$m, where the proposed methods demonstrate lower RMSE.

In terms of outliers (Fig.~\ref{fig: outliers}, middle plot), AR reported 64$\%$, 25.5$\%$ and 7$\%$ for $d \!=\! 1$~$\mu$m, $d \!=\! 1.5$~$\mu$m and $d \!=\! 2$ $\mu$m, respectively. Hankel methods showed reduced percentages at the same sample thicknesses: 60.5$\%$, 24$\%$, and 6$\%$ for HK and 59.5$\%$, 24$\%$, and 6$\%$ for RHK. For $d \!=\! 2.5$ $\mu$m, AR reported 8$\%$ outliers, while the proposed approaches achieved zero outliers.

The third row of Fig.~\ref{fig: estimation} demonstrates that as $\alpha$ increases, the RMSE slightly increases for the estimation of $c$ and $\alpha$. {All methods yield similar RMSEs, with slightly higher errors observed for AR in some cases. In addition, a small gap between the RMSE and the CRB suggests that there is still room for improvement in estimation accuracy.} Notably, no outliers were detected in these experiments, thus the corresponding outliers report is omitted from Fig.~\ref{fig: outliers}.

The bottom row in Fig.~\ref{fig: estimation} illustrates the impact of the bandwidth on the estimation performance. It is important to note that the CRB does not depend on the bandwidth, remaining constant across all scenarios. However, the gap between the RMSE and the CRB becomes more pronounced when smaller bandwidths are used. This can be explained due to a smaller bandwidth reduces the number of frequencies available for analysis, resulting in less data and, consequently, a more challenging estimation. In these cases, Hankel methods exhibited better performance than AR, while from a bandwidth of -10 dB, the performance of all estimators stabilizes and remains asymptotically close to the CRB. 

Outliers were observed only for bandwidths limited to -4 dB, -5 dB, and -6 dB, as shown in Fig.~\ref{fig: outliers}. HK and RHK exhibited a reduced number of outliers, reporting 14$\%$, 1$\%$, and 1$\%$, respectively, compared to AR, which reported 15.5$\%$, 11$\%$, and 16$\%$ outliers under the same conditions. These results further validate the choice of a bandwidth of -12 dB as appropriate for achieving accurate parameter estimation in the simulated data.

To summarize, CRB curves play an important role, as expected, in predicting the behavior of the estimation algorithms and their dependence on parameter values and noise levels. Specifically, CRB$_c$ depends on all parameters, while CRB$_Z$ is influenced by $Z$ and $d$. Additionally, CRB$_\alpha$ is dependent on all parameters except for $Z$, whereas CRB$_d$ depends on all parameters except for $\alpha$. Under favorable conditions, including high SNR, large $Z$ and $d$, and low $\alpha$, all methods show accurate performance with RMSE values approaching the CRB, demonstrating the reliability of the estimators. In challenging scenarios such as low SNR, small impedance contrasts, and thin samples, the proposed Hankel methods, particularly RHK, consistently outperform AR, achieving comparable or better estimation performance and significantly reduced number of outliers.

\subsection{Experiments}

\begin{figure}
\centering
\includegraphics[width=1\textwidth,trim=30 20 30 20, clip]{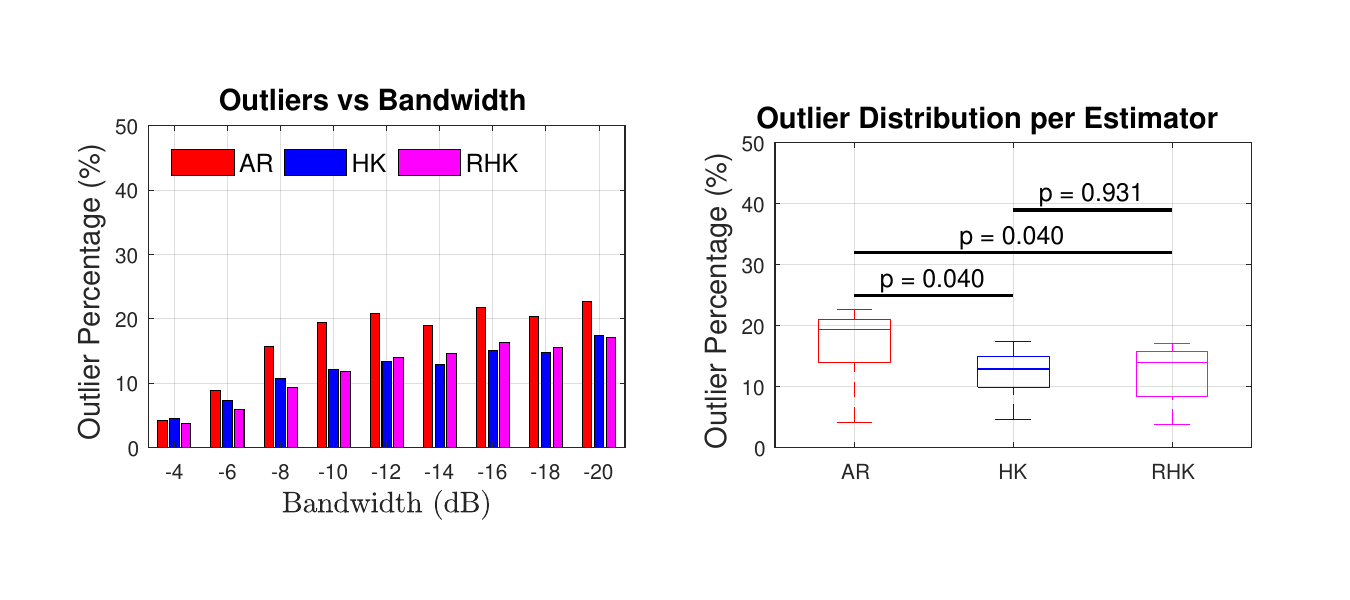}
\caption{{Outlier analysis on the lymph node ROI. Left: Percentage of outliers detected across varying bandwidths (from -4 to -20 dB) for AR (red), HK (blue), and RHK (magenta). AR consistently yields higher outlier rates, particularly at wider bandwidths. {Right:} Boxplots showing the distribution of outlier percentages for each method. Reported $p$-values are from the Wilcoxon rank-sum test.}}
\label{fig: out exp}
\vspace{-3mm}
\end{figure}

We conducted an experiment to assess how bandwidth selection affects the occurrence of unreliable estimates within the lymph node ROI. In Fig.~\ref{fig: out exp}, the left plot shows the percentage of outliers detected for bandwidths ranging from -4 to -20 dB. The right plot presents boxplots summarizing the distribution of outlier rates for each method across this range.

All methods exhibited sensitivity to bandwidth, with outlier rates generally increasing as the bandwidth expanded. {To determine whether these differences were statistically significant, we performed pairwise comparisons using the non-parametric Wilcoxon rank-sum test \cite{Fay2010}. This test was selected due to its robustness to non-normal distributions and its focus on median differences. The AR estimator showed significantly higher outlier rates compared to both HK and RHK, with $p$-values of 0.040 in both cases. No significant difference was observed between HK and RHK ($p = 0.931$). These results suggest that Hankel methods are more robust than AR under experimental conditions.}

Interestingly, this results contrasts with the simulation results, where reducing the number of frequency points (narrowing bandwidth) typically led to poorer performance. In real data, however, wider bandwidths include frequency components farther from the central frequency, which tend to introduce non-stationary behavior and additional noise, such as electronic interference, not accounted for in the current model.

\begin{figure*}[tb]
    \centering
    \includegraphics[width=1\textwidth,trim=0 0 100 0, clip]{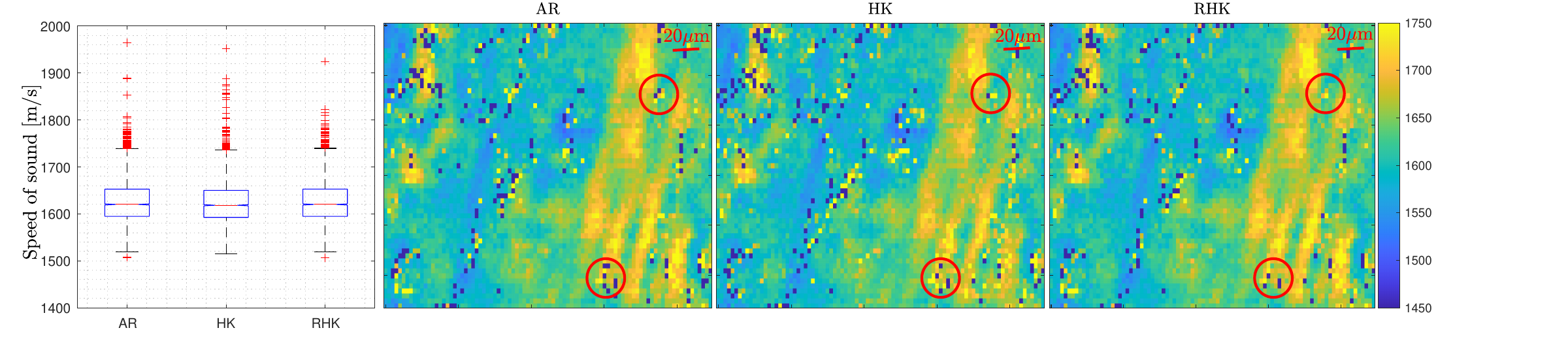}
    \includegraphics[width=1\textwidth,trim=0 0 100 10, clip]{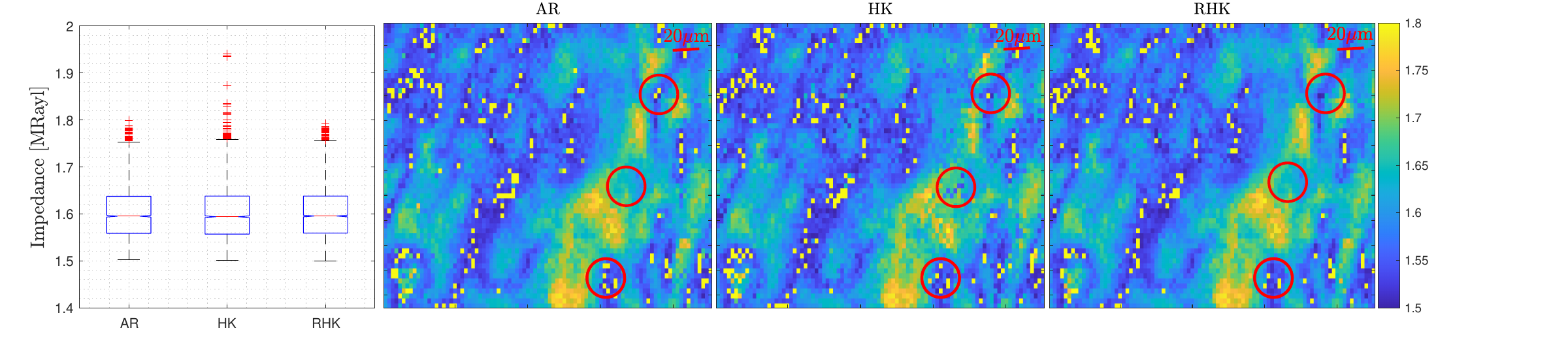}
    \includegraphics[width=1\textwidth,trim=0 0 100 10, clip]{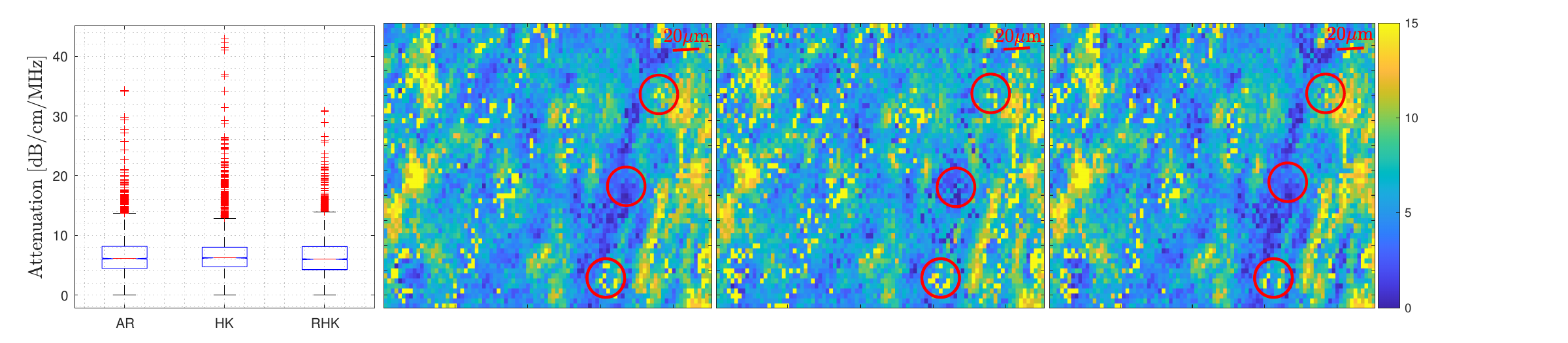}
    
    \caption{Estimation results in real data ROI for a bandwidth at -4 dB (AR: 6$\%$, HK: 5$\%$, RHK: 4$\%$ outliers). Column 1: Box plot illustrating the distribution of parameter estimates after outlier removal for improved visualization. Columns 2-4: Results for AR, HK, and RHK methods. Rows 1-3: Correspond to $c$, $Z$ and $\alpha$, respectively.}
    \label{fig:exp}
    \vspace{-3mm}
\end{figure*}

Fig.~\ref{fig:exp} presents the estimation results for a bandwidth at -4 dB, which reported the lowest outlier rates across all methods (AR: 6$\%$, HK: 5$\%$, RHK: 4$\%$). The first column displays box plots illustrating the distribution of parameter estimates after outlier removal for improved visualization. For all parameters, the median (red line) and interquartile range (blue box width) are similar across all algorithms. Notably, RHK consistently produces narrower distributions compared to HK and AR, reflecting improved stability and robustness. 

\begin{figure}[tb]
    \centering
    \includegraphics[width=1\textwidth,trim=0 0 0 0, clip]{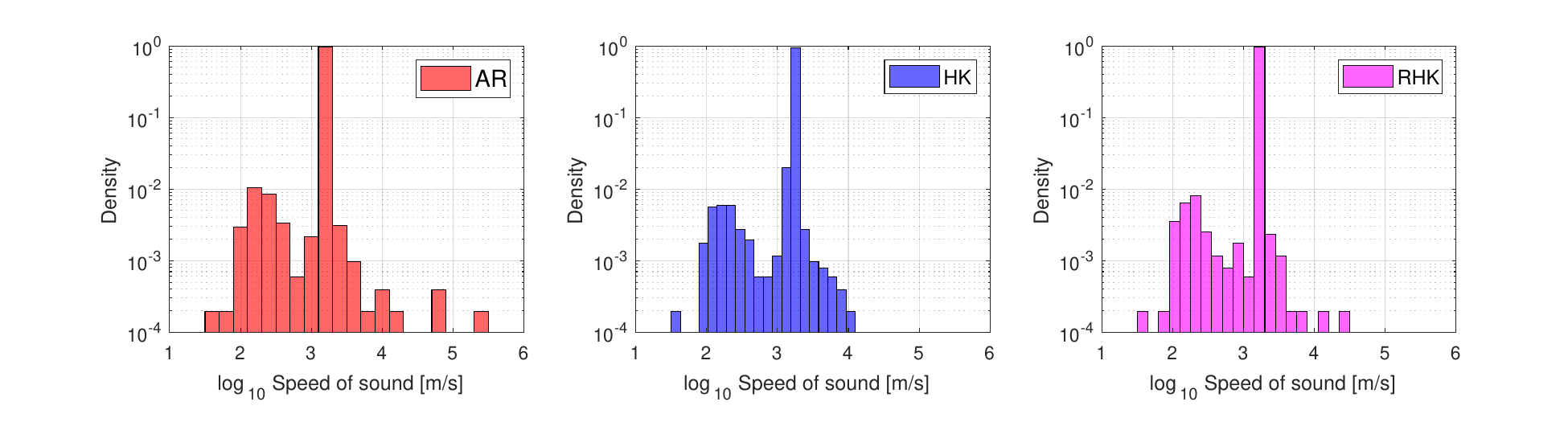}
    \includegraphics[width=1\textwidth,trim=0 0 0 0, clip]{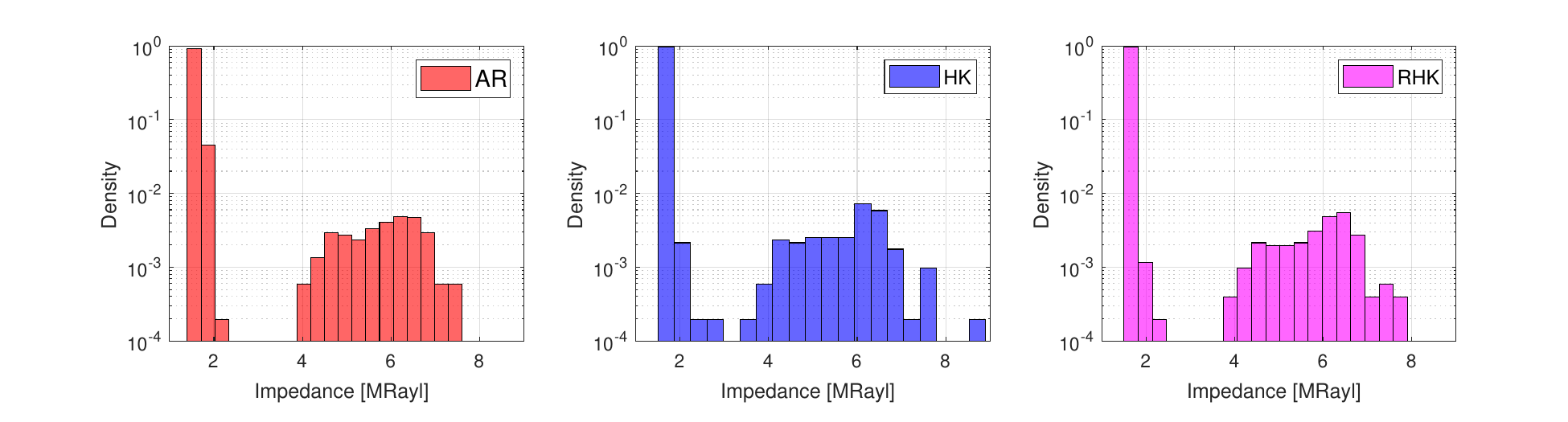}
    \includegraphics[width=1\textwidth,trim=0 0 0 0, clip]{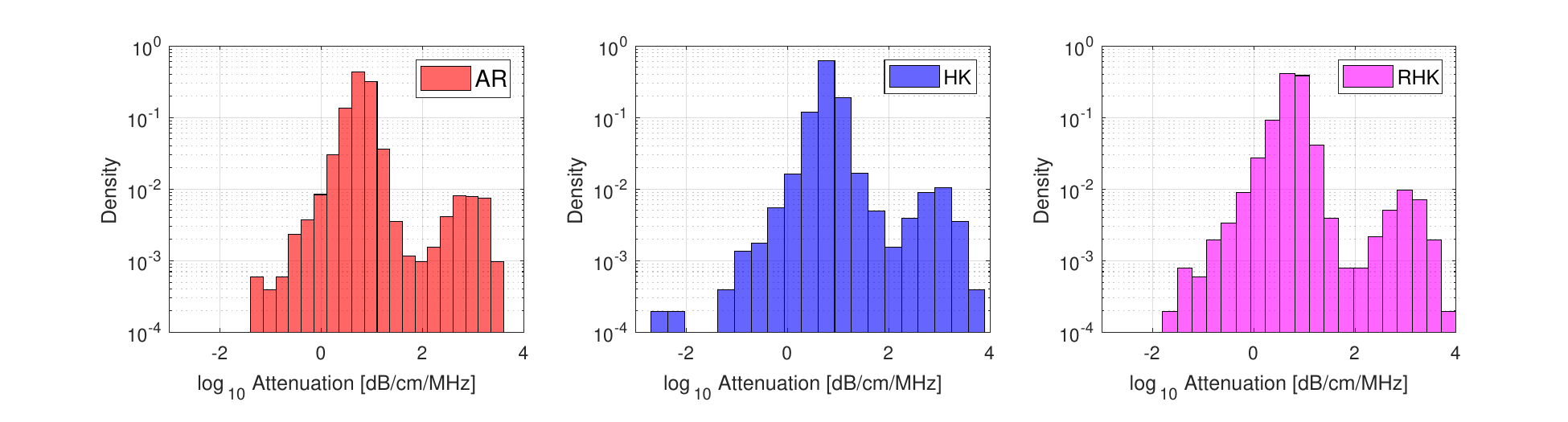}
        \caption{Distribution densities of estimated parameters ($c$, $Z$, and $\alpha$), including outliers, from the lymph node ROI. Each row corresponds to one parameter across the three methods (AR, HK, RHK).}
    \label{fig: histograms}
       \vspace{-3mm}
\end{figure}

Columns 2 to 4 show the corresponding estimation maps without outlier removal for AR, HK, and RHK methods. The rows represent $c$, $Z$ and $\alpha$, respectively. Pixels with  small (dark blue) or large (intense yellow) unreliable values are more prevalent in AR and HK than RHK, consistent with their slightly larger outlier rates. Regions highlighted by red circles illustrate the visual improvements offered by RHK. {While the visual differences may appear modest, this is primarily because the selected case represents a relatively favorable condition in which all methods, including AR, perform reasonably well. The advantages of the Hankel-based approaches are better demonstrated by their robustness in more challenging scenarios, as shown in Fig.~\ref{fig: out exp}. Nevertheless, we emphasize that even modest improvements are meaningful in the context of QAM, given the complexity of high-frequency ultrasound data and the inherent challenges of robust quantitative parameter estimation. Such advances support the broader goal of making QAM technology more accessible and practical.}

{As a complementary analysis, we present in Fig.~\ref{fig: histograms} the true distribution densities of the estimated parameters, including outliers, for the lymph node ROI. To assess whether the estimation methods produce significantly different distributions, we conducted pairwise two-sample Kolmogorov-Smirnov  tests~\cite{Simard2011}, which are well suited for detecting differences between potentially multimodal distributions. The results, shown in Table~\ref{tab:ks_tests}, reveal that for attenuation, all comparisons between AR, HK, and RHK yielded statistically significant differences ($p<0.05$), indicating that the Hankel methods produce distributions distinct from AR. For speed of sound and impedance, however, no statistically significant differences were found.}

\begin{table}[tb]
   \vspace{-3mm}
\centering
\caption{Results of two-sample Kolmogorov-Smirnov tests comparing the distributions of estimated parameters across methods. Statistically significant differences ($p < 0.05$) are shown in bold.}
\label{tab:ks_tests}
\begin{tabular}{lccc}
\toprule
\textbf{Parameter} & \textbf{Comparison} & \textbf{p-value} & \textbf{KS statistic} \\
\midrule
\multirow{3}{*}{Attenuation $\alpha$} 
& AR vs. HK   & \textbf{$<$0.0001} & 0.0458 \\
& AR vs. RHK  & \textbf{0.0259}    & 0.0290 \\
& HK vs. RHK  & \textbf{$<$0.0001} & 0.0713 \\
\midrule
\multirow{3}{*}{Speed of Sound $c$} 
& AR vs. HK   & 0.1639  & 0.0220 \\
& AR vs. RHK  & 1.0000  & 0.0060 \\
& HK vs. RHK  & 0.1433  & 0.0226 \\
\midrule
\multirow{3}{*}{Impedance $Z$} 
& AR vs. HK   & 0.0694  & 0.0255 \\
& AR vs. RHK  & 0.9991  & 0.0073 \\
& HK vs. RHK  & 0.0638  & 0.0259 \\
\bottomrule
\end{tabular}
\end{table}

While the computational complexity of the frequency estimation algorithms was discussed in Section~\ref{sec: adm}, we provide here the computational runtime for the experimental dataset as an illustrative example. On standard hardware (Intel i7, 32 GB RAM), processing a $60 \times 90$ pixel region takes approximately 3 minutes for HK and 8 minutes for RHK, compared to under 1 minute for AR and ESPRIT. \REV{Although runtime is not a critical constraint in QAM applications, since processing is typically performed offline, accelerating QAM image formation remains an important open direction, though beyond the scope of this work. Potential avenues include adopting techniques developed for fast low-rank matrix completion in other imaging modalities, such as SVD-free solvers~\cite{Guo2017,kim2018loraks}, ADMM-free optimization methods~\cite{Ongie2017,Zhao2021}, and deep unfolding strategies~\cite{Huang2023}.}

\begin{figure}
\centering
\includegraphics[width=1\textwidth, trim=0 40 300 30,clip]{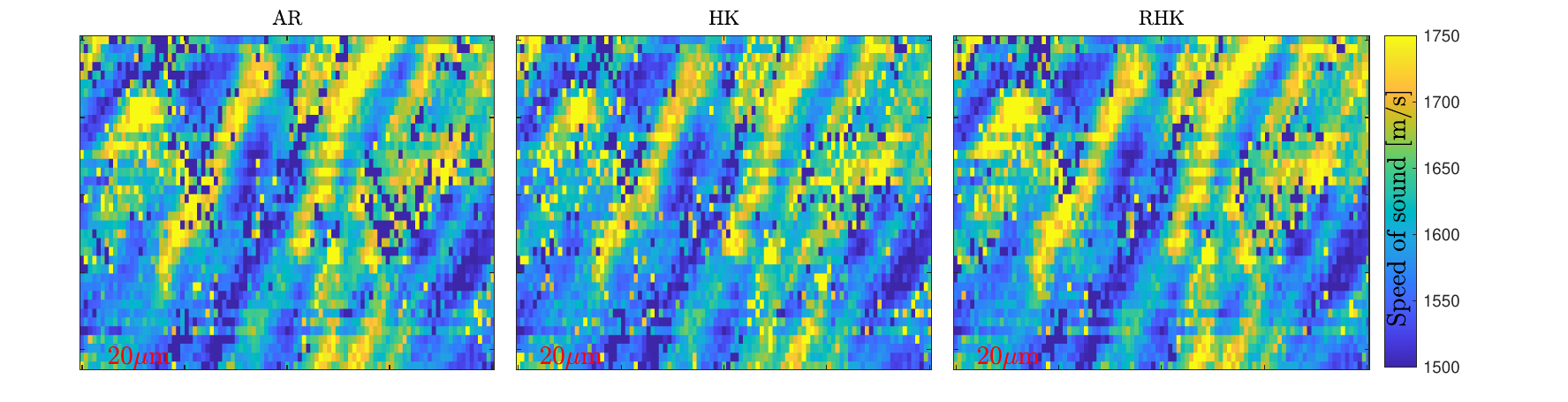}
\includegraphics[width=1\textwidth, trim=0 40 300 30,clip]{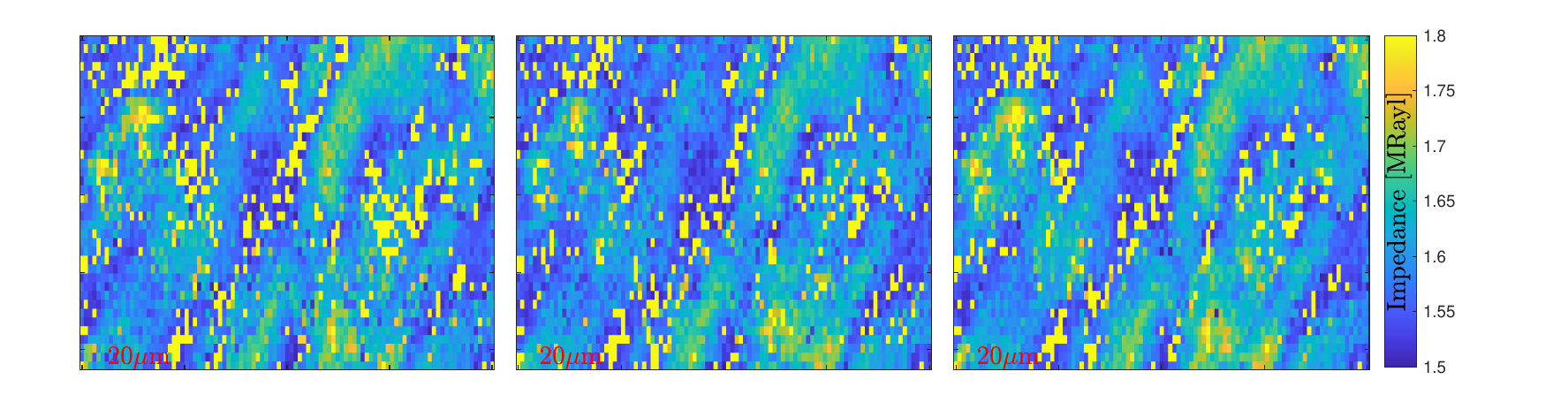}
\includegraphics[width=1\textwidth, trim=0 40 300 30,clip]{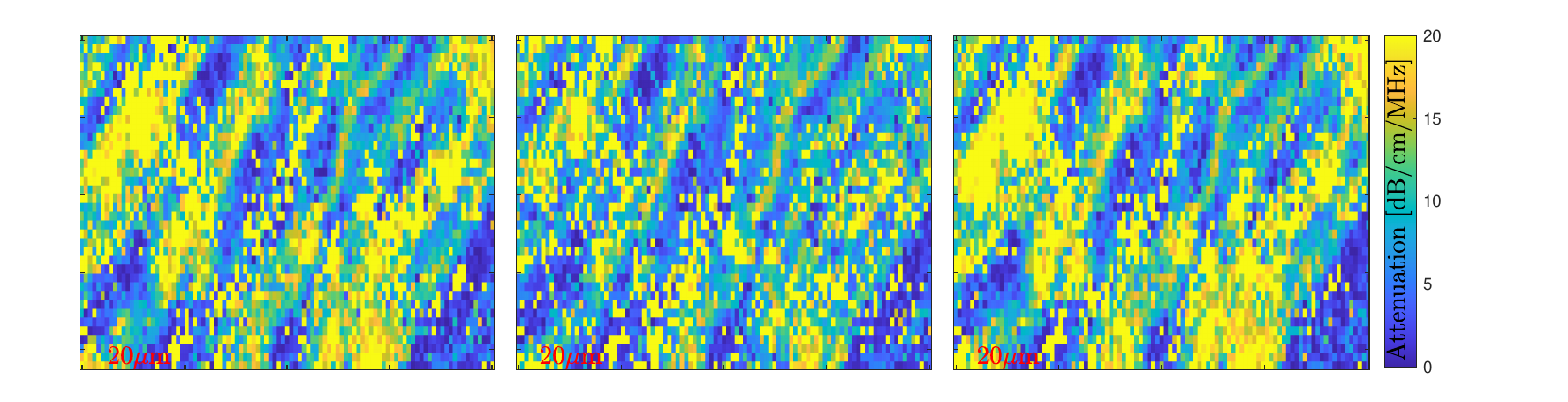}
\caption{Estimation results in the pig cornea ROI for a bandwidth threshold of $-4$~dB. Each row corresponds to $c$, $Z$, and $\alpha$, respectively, for the AR, HK, and RHK methods. Reported outlier rates: AR = 12$\%$, HK = 10$\%$, RHK = 9$\%$.}
\label{fig: cornea}
\end{figure}

{Similar results were observed in the selected pig cornea ROI. RHK yielded a slightly lower number of outliers across bandwidths compared to HK, while AR consistently reported higher outlier rates in all cases. Fig.~\ref{fig: cornea} shows the acoustic parameter estimation maps for a bandwidth threshold of -4~dB, where the lowest outlier rates were obtained: AR = 12\%, HK = 10\%, and RHK = 9\%. Some visual improvements can be observed with RHK, particularly in the estimation of $c$ and $Z$, compared to the other methods.}

To mitigate outliers, standard QAM post-processing image enhancement techniques can be applied to the estimated parameter maps. However, these techniques were omitted in this study to ensure a fair comparison between the different estimation frameworks.

\section{Conclusions and Perspectives}\label{sec:conc}

In this work, we proposed a weighted Hankel framework for high-frequency signal processing in QAM. This method extends our previous approach by integrating a reweighting strategy based on redescending M-estimators. For the first time in the context of QAM, we derived the Cramér-Rao bounds, providing theoretical performance benchmarks for analyzing the accuracy of acoustic parameter estimation. 
Results in both simulated and experimental data at 500~MHz, demonstrated that the proposed approach consistently reduced outliers and while keep or improve parameter estimation accuracy compared to the standard AR approach, especially in challenging scenarios involving high noise levels or low impedance contrasts. These advancements come at a moderate computational cost, making the RHK framework a practical and effective alternative for real-world QAM applications.

{Future work will seek to extend the scope and efficiency of Hankel frameworks for QAM imaging. The approaches will be validated at high noise levels and different transducer frequencies, such as advanced 1~GHz QAM systems. We are also investigating strategies to accelerate the methods, with the goal of enabling faster QAM imaging. In addition, we plan to investigate joint estimation techniques that take into account spatial correlations in tissue, which may lead to more coherent and accurate parametric maps.}




\section{Compliance with Ethical Standards}
The protocol (22-06024887) was approved by Weill Cornell Medicine Institutional Review board on August 2, 2023.

\section{Acknowledgments}
We are grateful to the authors of \cite{Andersson2014} who have kindly provided us with the original code of their paper. This work was
supported by the National Institutes of Health (NIH) grant R01GM143388 (JM).

\bibliographystyle{elsarticle-num} 
\bibliography{refs}






\end{document}